\documentclass[a4paper,twocolumn,11pt,accepted=2026-11-16]{quantumarticle}
\pdfoutput=1
\usepackage[utf8]{inputenc}
\usepackage[english]{babel}
\usepackage[T1]{fontenc}
\usepackage{amsmath}
\usepackage{amssymb}
\usepackage{amsthm}
\usepackage{bbm}
\usepackage{braket}
\usepackage{graphicx}
\usepackage{xcolor}
\usepackage{hyperref}

\definecolor{riverlane_green}{RGB}{0, 111, 98}
\definecolor{riverlane_light_green}{RGB}{0, 150, 143}
\definecolor{riverlane_orange}{RGB}{255, 117, 0}
\definecolor{riverlane_red}{RGB}{220, 68, 5}
\definecolor{riverlane_pink}{RGB}{207, 111, 127}
\hypersetup{
  colorlinks   = true, %Colours links instead of ugly boxes
  urlcolor     = riverlane_green, %Colour for external hyperlinks
  linkcolor    = riverlane_orange, %Colour of internal links
  citecolor   = riverlane_green  %Colour of citations
}

\newcommand{\OVTV}{O_{\mathrm{VTV}}}
\newcommand{\OVTT}{O_{\mathrm{VTT}}}
\newcommand{\abs}{\mathrm{abs}}
\newcommand{\NR}{N_{\mathrm{R}}}
\newcommand{\NT}{N_{\mathrm{T}}}
\newcommand{\one}{\mathbbm{1}}

\newtheorem{theorem}{Theorem}[section]
\newtheorem{lemma}[theorem]{Lemma}

\begin{document}

\title{Quantum simulation of nanographenes and Trotter error cancellation}

\author{Andreas Juul Bay-Smidt}
\email{andreas.bay-smidt@nbi.ku.dk}
\affiliation{NNF Quantum Computing Programme, Niels Bohr Institute, University of Copenhagen, Denmark}
\affiliation{Nano-Science Center and Department of Chemistry, University of Copenhagen, Denmark}
\author{Nina Glaser}
\affiliation{NNF Quantum Computing Programme, Niels Bohr Institute, University of Copenhagen, Denmark}
\affiliation{Nano-Science Center and Department of Chemistry, University of Copenhagen, Denmark}
\author{Marcel D. Fabian}
\affiliation{NNF Quantum Computing Programme, Niels Bohr Institute, University of Copenhagen, Denmark}
\author{Earl T. Campbell}
\affiliation{Riverlane, Cambridge, CB2 3BZ, UK}
\affiliation{School of Mathematical and Physical Sciences, University of Sheffield, Sheffield, S3 7RH, UK}
\author{Nick S. Blunt}
\email{nick.blunt@riverlane.com}
\affiliation{Riverlane, Cambridge, CB2 3BZ, UK}
\author{Gemma C. Solomon}
\email{g.solomon@chem.ku.dk}
\affiliation{NNF Quantum Computing Programme, Niels Bohr Institute, University of Copenhagen, Denmark}
\affiliation{Nano-Science Center and Department of Chemistry, University of Copenhagen, Denmark}

\maketitle

\begin{abstract}
Fault-tolerant quantum computing is a promising tool for simulating molecules and materials, but frequently-considered applications require substantial resources, and the gap between hardware capabilities and requirements remains significant. We propose quantum simulation of nanographene $\pi$-systems as relevant and scalable problems to span the gap between early and large-scale fault-tolerant quantum computing. We examine the efficiency of Trotterized quantum simulation, present a detailed analysis of worst-case, average-case and energy eigenvalue Trotter errors, and show that these Trotter error estimates vary by orders of magnitude. Trotter eigenvalue errors are obtained from a novel tensor-network-based approach which allows spectral analysis of product formulas for systems beyond brute-force calculation. Notably, we observe a Trotter error cancellation phenomenon whereby the Trotter error for energy differences between low-lying eigenstates is significantly smaller than the Trotter error for absolute energies, resulting in approximately an order of magnitude circuit depth reduction for quantum phase estimation calculation of energy gaps. This is a significant result because for most chemical applications, only energy differences are of practical relevance. We estimate that calculation of energy gaps to chemical accuracy between the ground- and excited-states within the Pariser--Parr--Pople model for large 2D nanographenes (up to 140 spin orbitals) requires circuits with $< 3.2 \times 10^7$ Toffoli gates. This work shows that considering details of chemically-relevant applications and exploiting error cancellation can lead to substantial reductions in resource requirements.
\end{abstract}

\section{Introduction \label{sec:introduction}}
Electronic structure problems in chemistry and materials science are generally too complex to be solved exactly. Nonetheless, approximate (classical) computational methods play an important role in explaining experimental results and delivering theoretical understanding. Two classes of approximations are typically made.

First, simulations do not exactly mimic experiments; rather, relevant quantities are calculated in order to explain physical observables. For example, reaction rates are typically predicted on the basis of energy differences between reactants, transition states, and products. Similarly, energy gaps between electronic states are used to describe photochemical and photophysical properties, and energy differences along potential energy surfaces are used to perform dynamics of molecular systems.

Second, approximate (classical) computational methods are used to solve electronic structure and dynamics problems. The accuracy of a method does not predict how widely it is used, as the cost of a given method versus the nature of the scientific question or the system size often dictate choices. For example, variants of density functional theory \cite{Hohenberg1964, Kohn1965} are widely used, despite shortcomings such as delocalization and static-correlation errors \cite{Cohen2008}, due to their efficiency and ease of use \cite{Burke2012, Mardirossian2017}. On the other hand, wavefunction methods for handling strong correlation, such as quantum Monte Carlo \cite{Foulkes2001, becca_sorella_2017} and the density matrix renormalization group (DMRG) algorithm \cite{White1992, Amaya2015}, trade improved accuracy for higher computational complexity.

In practice, the success of approximate methods often relies on some degree of error cancellation, making the performance of these methods exceed what could be reasonably expected \cite{Helgaker2004, Scemama2018FNDMExcitations, cances2017}, for example for calculating chemically-relevant quantities like energy differences. The approximations to solve a given problem are also important for pragmatic reasons to best utilize available resources. The same will apply to quantum computers, especially in the early fault-tolerant era.

Quantum computing provides a new set of computational tools that may circumvent limitations of classical methods with respect to accuracy guarantees and problem complexity \cite{McArdle2020}. Simulation of quantum many-body systems using quantum computers, commonly referred to as quantum simulation, is therefore considered an interesting direction towards a scalable method for quantum chemistry problems with accuracy guarantees. Eigenvalue estimation and quantum dynamics are two key applications of quantum simulation, both of which can be achieved by implementing Hamiltonian simulation $\mathcal{U} = e^{-iHt}$ \cite{Abrams1997, Kassal2008Polynomial}, which we also refer to as the time evolution operator. Trotterization (or ``product formulas'') is a conceptually simple quantum algorithm to approximately implement Hamiltonian simulation \cite{Trotter1959OnOperators, Suzuki1991GeneralPhysics, Lloyd1996UniversalSimulators}. The error associated with Trotterization is determined by the difference between the exact Hamiltonian simulation operator $\mathcal{U}$, and its Trotterized approximation $U$. The error is dependent on the choice of time step, $t$, as well as on constant factors \cite{Childs2021TheoryScaling}. Well-established worst-case Trotter error analysis of these constant factors can lead to overestimation of errors for certain applications. For example, average-case Trotter errors better predict true simulation errors for certain quantum dynamics simulations \cite{Zhao2022, Zhao2024l}. Eigenvalue errors from Trotterized quantum phase estimation (QPE) are typically also overestimated by worst-case errors \cite{Sahinoglu2021, gunther2025phaseestimationpartiallyrandomized}, and it is more appropriate to consider differences between the eigenvalues of $H$ and those of the effective Hamiltonian $\tilde{H}$, which generates the Trotterized time-evolution, $U = e^{-i \tilde{H}t}$ \cite{Reiher2017Elucidating, Yi2022, gunther2025phaseestimationpartiallyrandomized}. The error between $H$ and $\tilde{H}$ can be controlled by decreasing $t$, which in turn increases the total simulation cost for QPE.

Trotterized quantum simulation of Hubbard models is a commonly-considered early fault-tolerant quantum computing application \cite{Campbell_2022, Bay-Smidt2025, Kan2025, Toshi2025, Akahoshi2025, Chung2026} due to a combination of low qubit requirements and sparsity of the model. On the other hand, \emph{ab initio} chemistry problems considered for fault-tolerant quantum computing are typically complex, require substantial resources, and these resources are typically estimated with reference to calculating single-point (absolute) energies, which are usually not the most chemically-relevant quantities. This suggests a lack of chemically-relevant problems that span the gap between suitable early fault-tolerant and large-scale fault-tolerant problems and which also directly address how to extract relevant quantities to solve the problem of interest.

We address these gaps in two steps. First, we present a set of relevant problems involving nanographenes modeled by the Pariser--Parr--Pople (PPP) model, which is essentially a Hubbard model with all-to-all interactions and therefore a natural step beyond the Hubbard model towards \emph{ab initio} chemistry. We highlight examples of problems where vertical excitation energies, or ``energy gaps'', are the primary quantity of interest. Second, we relate these problems directly to quantum resource requirements and show that considering the details of chemically-relevant applications can significantly reduce quantum simulation costs. This is achieved through a detailed Trotter error analysis. In particular, we study the worst-case, average-case and energy eigenvalue Trotter errors for a range of nanographenes. The Trotter error in energy eigenvalues is calculated by a novel analysis combining tensor network simulation with time series analysis, allowing us to perform spectral analysis of product formulas to larger system sizes, well beyond the reach of approaches based on exact construction and diagonalization. Crucially, we show that the Trotter error on low-lying energy gaps is significantly smaller than the Trotter error on absolute energies. That is, \emph{there is a significant degree of Trotter error cancellation} between low-lying energy eigenvalues. We furthermore estimate that the observed Trotter error cancellation reduces circuit depth requirements of quantum phase estimation for calculating energy gaps by approximately an order of magnitude compared to calculating absolute energies. This is an important result because energy differences are usually the most chemically-relevant quantity.

In summary, in Section~\ref{sec:nanographenes} we introduce the structure, properties and applications of nanographenes whose chemical and physical properties are governed by their $\pi$-systems \cite{Clar1983, Suresh1999, Gu2022Nanographenes, fabian2025pppmodelminimal}. We argue that simulation of nanographenes provides a set of relevant and scalable problems with a natural path toward classically challenging regimes. In Section~\ref{sec:PPP_model}, we present the PPP model, an effective model of nanographene $\pi$-systems. In Sections~\ref{sec:ham_sim_and_Trotter_error} and~\ref{sec:trotter_error_results}, we examine the efficiency of Trotterized quantum simulation of nanographenes by comparing worst-case, average-case, energy and gap errors of second-order split-operator (SO) Trotter schemes \cite{Kivlichan2020ImprovedTrotterization, Campbell_2022, Bay-Smidt2025}. Worst- and average-case errors can be evaluated via commutator bounds \cite{Childs2021TheoryScaling, Zhao2022}. We evaluate the worst-case SO-error using a Monte Carlo method \cite{blunt2025montecarloapproachbound}, and establish tighter error bounds on free-fermionic operator splittings as used in Refs.~\cite{Campbell_2022, Bay-Smidt2025}. Energy and gap errors are errors on eigenenergies and eigenenergy differences between the true and the effective Hamiltonian. The four error types vary by orders of magnitude for the systems considered here, highlighting the importance of application-specific Trotter error analysis. Using our results on Trotter error cancellation, we show that it is possible to use Trotter product formulas with large time step sizes for energy gap calculations to chemical accuracy using QPE. In Section~\ref{sec:resource_estimates}, we compare the resulting QPE non-Clifford resource estimates for nanographenes using the four different Trotter errors, and demonstrate that targeting energy gaps can reduce circuit depth and non-Clifford costs by approximately an order of magnitude.

\section{Nanographenes \label{sec:nanographenes}}
\begin{figure*}[t]
    \centering
    \includegraphics[width=0.82\linewidth]{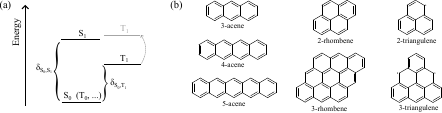}
    \caption{(a) Illustration of quantities of interest of nanographenes that we consider in this paper. This includes energy gaps between a ground state singlet ($\mathrm{S_0}$) and a first-excited triplet ($\mathrm{T_1}$) or a first excited singlet ($\mathrm{S_1}$), the energy of $\mathrm{T_1}$ compared to $\mathrm{S_1}$ (where the gray arrow and additional $\mathrm{T_1}$ energy line indicate that in certain instances $E_{\mathrm{T_1}} \gtrsim E_{\mathrm{S_1}}$), and the character of the ground state, e.g. whether the ground state is $\mathrm{S_0}$, $\mathrm{T_0}$ or another high-spin state. (b) Three different classes of nanographenes: (left) $n$-acenes, where $n$ denotes the number of hexagons, (center) $n$-rhombenes, where $n$ denotes the number of hexagons along the edges, and (right) $n$-triangulenes, where $n$ denotes the number of hexagons along each side of the equilateral triangulene. Each corner of a hexagon represents a carbon atom, and using the size parameter, $n$, we can calculate the number of carbon atoms: $N_{\mathrm{acene}} = 6 + 4(n-1)$, $N_{\mathrm{rhombene}} = 2(n+1)^2-2$ and $N_{\mathrm{triangulene}} = n^2 + 4n +1$. }
    \label{fig:molecular_structures_nanographenes}
\end{figure*}

Nanographenes, or polycyclic aromatic hydrocarbons (PAHs), consist of finite fragments of a hexagonal lattice of carbon atoms with hydrogen atoms at the perimeter, and they exhibit a wide range of interesting and chemically-tunable properties \cite{Gu2022Nanographenes, Dai18042018, Narita2015}. One of the unifying characteristics of nanographenes is what is known as their $\pi$-system, which is defined as the electrons that occupy a subset of orbitals that are antisymmetric to the plane of the lattice. The $\pi$-system governs many of their physical and chemical properties \cite{fabian2025pppmodelminimal}, including magnetic \cite{song2023highlyentangledpolyradicalnanographenecoexisting, Du2023OrbitalSymmetry}, opto-electronic \cite{Zirzlmeier2015, Narita2015} and semiconducting \cite{Mullen2008} properties. These properties are largely determined by the character of and energy gaps between low-lying electronic states. We briefly outline three physical phenomena, illustrated by the energy diagram in Fig.~\ref{fig:molecular_structures_nanographenes}(a), and potential applications of nanographenes and nanographene derivatives that originate from their low-lying electronic states to motivate practical interest in the systems considered here.

\textbf{Singlet fission} is a photo-physical process of high interest to applications in photovoltaics \cite{Hanna2006_carriermultiplication, Smith2010_singletfission}. This process can occur between molecules with a singlet ground state ($\mathrm{S_0}$), first-excited triplet ($\mathrm{T_1}$) and first-excited singlet ($\mathrm{S_1}$), as illustrated in Fig.~\ref{fig:molecular_structures_nanographenes}(a), where the energy differences $\delta$ between the electronic states fulfill $\delta_{\mathrm{S_0},\mathrm{S_1}} \geq 2 \delta_{\mathrm{S_0},\mathrm{T_1}}$ \cite{Zirzlmeier2015, Zimmerman2010_singletfission}.

\textbf{Inverted singlet-triplet energy gap} (InveST) \cite{de_silva_inverted_2019, perezjimenez_role_2025} is a phenomenon where $\mathrm{S_1}$ is lower in energy than $\mathrm{T_1}$, $E_{\mathrm{S_1}} < E_{\mathrm{T_1}}$, which allows for the development of more efficient organic light emitting diodes (OLEDs) \cite{POLLICE20211654, Aizawa2022}. This is illustrated in Fig.~\ref{fig:molecular_structures_nanographenes}(a) as finding the relative placement of $\mathrm{T}_1$ compared to $\mathrm{S}_1$, for example by calculating the energies $E_{\mathrm{S}_1}$ and $E_{\mathrm{T}_1}$, or the energy differences between states.

\textbf{Magnetic properties} and high spin-spin correlation arise from strongly interacting electrons and topological frustration which can lead to antiferromagnetically- or ferromagnetically-coupled spins in the ground state of nanographenes \cite{song2023highlyentangledpolyradicalnanographenecoexisting}. Designing stable molecules with certain ground-state spin properties (e.g. $\mathrm{S_0}$, $\mathrm{T_0}$ or other high-spin states) has applications in spintronics and for high-spin materials \cite{Gu2022Nanographenes, Du2023OrbitalSymmetry}.

Fig.~\ref{fig:molecular_structures_nanographenes}(b) shows the three classes of nanographenes considered in this work: acenes, rhombenes and triangulenes. Each corner of a hexagon represents a carbon atom, and each carbon atom contributes one electron to the $\pi$-system. The hydrogen atoms at the edges are omitted, and will not be considered in the remaining text since they do not contribute to the $\pi$-system. While these systems do not represent the full complexity landscape of nanographenes and PAHs, they cover a range of shapes, properties and applications, and provide links to other structures that allow the findings of this paper to be applied more generally. We provide three reasons for the interest in acenes, rhombenes and triangulenes:

\textbf{1) Chemical and physical properties:} Acenes, rhombenes and triangulenes have different geometries and Kekulé structures resulting in differences in their low-lying eigenspectra \cite{Das2016Polyradical}. Acenes have shown efficient singlet fission \cite{Zirzlmeier2015, Zimmerman2010_singletfission}, whereas rhombenes have strong antiferromagnetic coupling between electrons in their ground states \cite{Mishra2021_rhombene, Barragan2025_rhombene}. The ground state of $n$-triangulene is predicted to be high-spin with total spin $S=(n-1)/2$ due to the topology of these structures \cite{Borden1977, VilasVarela2023AzaTriangulene}. Nitrogen-doped triangulenes are promising candidates for OLED materials due to their inverted singlet-triplet gaps \cite{Bedogni2024, Aizawa2022}.

\textbf{2) Classical hardness and natural scalability:} The systems can be scaled by increasing $n$, leading to larger and more computationally challenging problems. The shapes of the nanographenes also provide a natural progression of classical hardness especially for standard tensor network algorithms, e.g. DMRG. The quasi-1D acenes are known to be DMRG-easy \cite{Hachmann2007}, while the 2D rhombenes and triangulenes are more challenging. The character of the $\pi$-system electronic structure problem can also be changed by exchanging carbon atoms with boron or nitrogen \cite{Hele2019AceneSpectra, Bedogni2024}.

\textbf{3) Possibility of verification:} These nanographenes have been, and continue to be, studied extensively both computationally and experimentally, which allows for verification of quantum simulation results. Experimental studies of new molecules and materials with atomic precision, for example using STM/AFM surface techniques \cite{song2023highlyentangledpolyradicalnanographenecoexisting, Pawlak2022, Su2021, Barragan2025_rhombene, Du2023OrbitalSymmetry}, give the possibility to simulate properties of novel and interesting chemical systems on the nanoscale and compare directly with experiment.

For these reasons, we believe that acenes, rhombenes and triangulenes represent a set of chemically relevant benchmark systems for fault-tolerant quantum computation, while also providing a natural pathway towards classically challenging problems.

\section{$\pi$-system model \label{sec:PPP_model}}
The nanographene $\pi$-systems are modeled with the semi-empirical Pariser--Parr--Pople (PPP) model \cite{Pariser1953, Pople1953, fabian2025pppmodelminimal} that contains nearest-neighbor hopping terms and all-to-all interaction terms. The PPP model has been shown to capture many physical properties of nanographenes, such as energy levels of their low-lying electronic states \cite{Chakraborty_2013, Lambie2025}, and optical \cite{sony_correlated_2005, Sony2007, chiappe_can_2015} and dynamical properties \cite{bhattacharyya_pariserparrpople_2020, Bostrom2018}.

It is convenient to group the PPP Hamiltonian terms into a kinetic energy ($T$) and a potential energy ($V$) operator, allowing us to define $\pi$-system models of nanographenes with $N$ carbon atoms (sites) and $2N$ spin orbitals as $H = T + V$ with
\begin{align}
     \! T &= - \tau \sum_{\langle ij \rangle, \sigma} a_{i \sigma}^{\dagger}a_{j \sigma}, \label{eq:T} \\
     \! V &= u \sum_i^N n_{i\uparrow}n_{i\downarrow} + \sum_{i<j} v_{ij} (n_{i} - \one) (n_{j} - \one), \label{eq:V}
\end{align}
where $a^\dagger_{i\sigma}$ and $a_{i\sigma}$ with $\sigma \in \{ \uparrow, \downarrow \}$ are fermionic creation and annihilation operators, $n_{i\sigma} = a^\dagger_{i\sigma} a_{i\sigma}$ and $n_{i} = n_{i\uparrow}+n_{i\downarrow}$. The potential energy term between electrons on sites $i$ and $j$ with intersite distance $r_{ij}$ is parameterized by the Ohno potential, $v_{ij} = u \big/ {\sqrt{1+\alpha r_{ij}^2}}$. Throughout this work, we use the following standard parameters $\tau = 2.4 \; \mathrm{eV}$, $u=11.13 \; \mathrm{eV}$, and $\alpha = 0.6117 \; \text{\AA}^{-2}$ \cite{Chakraborty_2013, Sony2007, Sony2009}, and fix the carbon-carbon bond length at $1.4$~\AA.

\begin{table}[t] \centering
\small \begin{tabular}{llll}
\hline
\hline
& & \multicolumn{2}{c}{ $\#$ terms} \\
\cline{3-4}
Molecule & $N$ \;\; & $V$ \; \; \; \; & $V'$ \\
\hline
3-acene & 14 & 406 & 290 \\
7-acene & 30 & 1830 & 1554 \\
\hline
3-rhombene & 30 & 1830 & 1522 \\
5-rhombene & 70 & 9870 & 8994 \\
\hline
3-triangulene & 22 & 990 & 778 \\
5-triangulene & 46 & 4278 & 3778  \\
\hline
\hline
\end{tabular}
\caption{Number of terms in the Pauli representation of the potential energy operator, $V$, and the shifted operator, $V'$, of PPP model nanographenes with $N$ carbon atoms.}
\label{tab:number_of_terms_V_and_V_shifted}
\end{table}

We implement Hamiltonian simulation using Trotterization whose cost per Trotter step scales explicitly with the number of terms in the Hamiltonian. The operators $T$ and $V$ contain $\mathcal{O}(N)$ and $\mathcal{O}(N^2)$ terms, respectively, such that the simulation cost is dominated by $V$ for sufficiently large $N$. We can reduce the cost by minimizing the number of terms in $H$, or more specifically in the Pauli-representation of $H$, obtained by a Jordan-Wigner (JW) transformation. We minimize the number of terms by performing symmetry shifts by operators that commute with $H$. Since we work within a fixed particle-number sector of the Hamiltonian \cite{Loaiza2023, MartinezMartinez2023assessmentofvarious}, we can shift $V$ by terms proportional to $\hat{N}$ and $\hat{N}^2$, where  $\hat{N}=\sum_i^{N} \sum_\sigma n_{i \sigma}$ is the number operator,
\begin{equation}
    V' = V + c_1 \hat{N} + c_2 \hat{N}^2,
\end{equation}
where $V'$ is the shifted potential energy operator, and $c_1$ and $c_2$ are adjustable parameters. This results in a constant energy shift of the eigenspectrum when fixing the number of electrons. In Appendix~\ref{app:sym_shift}, we outline a scheme for choosing $c_1$ and $c_2$ in order to most efficiently reduce the number of potential energy terms. Table~\ref{tab:number_of_terms_V_and_V_shifted} shows the effect of the symmetry shifts on a set of acenes, rhombenes and triangulenes. These shifts reduce the total cost of a Trotter step by approximately 10--25$\%$ for the systems studied here; while this does not dramatically reduce the cost of Trotterized quantum simulation, it provides an improvement at no additional cost.

\section{Hamiltonian simulation and errors \label{sec:ham_sim_and_Trotter_error}}
The applications considered here require us to implement Hamiltonian simulation of $H = T+V$. We consider approximate implementation using Trotterization, and we can evaluate the simulation error by comparing the exact Hamiltonian simulation unitary, $\mathcal{U} = e^{-iHt}$, to its Trotterized approximation, $U$. We consider two second-order Trotter schemes based on the split-operator (SO) formalism, where $T$ and $V$ are evolved separately
\begin{align}
    \! U_{\mathrm{SO}} &= e^{-iV \frac{t}{2}} e^{-iTt} e^{-iV \frac{t}{2}}, \label{eq:U_SO} \\
    \! U_{\mathrm{tile}} &= e^{-iV\frac{t}{2}} \prod_{s=1}^S e^{-iT_s \frac{t}{2}} \prod_{s=S}^1 e^{-iT_s \frac{t}{2}}  e^{-iV\frac{t}{2}}. \label{eq:U_SO_tile}
\end{align}
Here, $t$ denotes the time step of a single Trotter step, and all error measures below are defined per Trotter step. We discuss the total evolution time required for QPE in Section~\ref{sec:resource_estimates}. Compared to $U_{\mathrm{SO}}$ in Eq.~\eqref{eq:U_SO}, $U_{\mathrm{tile}}$ further decomposes $T$ into $S$ sections, $T=\sum_{s=1}^S T_s$, which trades a small additional Trotter error for reduced implementation cost to diagonalize the kinetic energy sections. Specifically, $U_{\mathrm{tile}}$ is implemented by the tile Trotterization approach of Ref.~\cite{Bay-Smidt2025}, which extended previous results of plaquette Trotterization in Ref.~\cite{Campbell_2022}. See Appendix~\ref{app:Trotter_errors} for more details.

We consider 1) worst-case, 2) average-case, 3) energy and 4) gap error, and associate a Trotter error constant to each. We will show that these errors differ by orders of magnitude so for a given application it is critical to consider which errors are most representative to accurately determine quantities of interest. We proceed to describe each error type and how it is computed (see Appendix~\ref{app:Trotter_errors} for details).

\textbf{1) Worst-case error} captures the worst-case state-vector difference between the exact unitary ($\mathcal{U}$) and its approximation ($U$), given by
\begin{equation}
\mathcal{W} (\mathcal{U} ,U) = \lVert \mathcal{U}  - U \rVert = \max_{\ket{\psi}} \lVert (\mathcal{U}  - U) \ket{\psi} \rVert_2,
\end{equation}
where $\lVert \, \cdot \, \rVert$ is the operator norm and $\lVert \, \cdot \, \rVert_2$ is the vector 2-norm (Euclidean norm). Worst-case error has been applied in both quantum dynamics and phase estimation \cite{Kivlichan2020ImprovedTrotterization, Campbell_2022, Bay-Smidt2025}, and typically overestimates actual simulation errors \cite{Zhao2024l, gunther2025phaseestimationpartiallyrandomized}, often by orders of magnitude. The worst-case error analysis is useful for asymptotic scaling considerations as well as for general quantum dynamics to strictly ensure accurate simulation (although it can overestimate both). The worst-case error of $U_{\mathrm{SO}}$ is bounded by $\lVert \mathcal{U}  - U_{\mathrm{SO}} \rVert \leq W_{\mathrm{SO}} t^3$, where $W_{\mathrm{SO}}$ is the worst-case Trotter error constant, which can be evaluated as \cite{suzuki_1985, Kivlichan2020ImprovedTrotterization}
\begin{equation}
     W_{\mathrm{SO}} = \frac{1}{24} \lVert [[V, T], V] \rVert + \frac{1}{12} \lVert [[V, T], T] \rVert.
     \label{eq:worst_case_so}
\end{equation}
The required spectral norms are widely upper bounded using the triangle inequality, which can significantly overestimate Eq.~\eqref{eq:worst_case_so}. Instead, we use the approach recently described in Ref.~\cite{blunt2025montecarloapproachbound}, which introduced a scalable numerical method based on Monte Carlo simulation. This method still results in an upper bound, but it was shown in Ref.~\cite{blunt2025montecarloapproachbound} that this is typically tighter than triangle-inequality-based results. The estimates of $\lVert [[V, T], V] \rVert$ in particular are essentially exact.

The Trotter error of $U_{\mathrm{tile}}$ can be upper bounded by adding $W_T$, a relatively small error contribution from the kinetic energy operator splitting, to $W_{\mathrm{SO}}$ \cite{Bay-Smidt2025}
\begin{equation}
     W_{\mathrm{tile}} \leq W_{\mathrm{SO}} + W_T.
\end{equation}
We describe in Appendix~\ref{app:free_fermionic_operators_error} how to obtain tighter bounds on $W_T$ than through nested-commutator approaches \cite{Campbell_2022, Bay-Smidt2025}.

\textbf{2) Average-case error} is of similar nature to the worst-case error, but state-vector errors are instead averaged over an ensemble of input states, $\ket{i}$, and upper-bounded by \cite{Zhao2022}
\begin{equation}
    \mathcal{A}(\mathcal{U},U) \leq \Big[ \mathbb{E}_{\ket{i} \in {\mathcal{E}}} \lVert (\mathcal{U} - U) \ket{i} \rVert_2^2 \Big ]^{1/2},
\end{equation}
where $\mathbb{E}$ estimates the average of $\lVert (\mathcal{U} - U) \ket{i} \rVert_2^2$ over an ensemble, $\mathcal{E}$, of states. We sample states from the simulation-relevant subspace of computational basis states at half-filling and for a given spin sector. The average-case error is an especially useful measure of Trotter error for systems with high entanglement entropy \cite{Zhao2024l}. The average-case error of $U_{\mathrm{SO}}$ is upper-bounded by $\mathcal{A}(\mathcal{U},U_{\mathrm{SO}}) \leq A_{\mathrm{SO}} t^3$  \cite{Zhao2022}, with
\begin{equation}
         \! A_{\mathrm{SO}} = \frac{1}{24 \sqrt{d_\mathcal{E}}} \big( \lVert [[V, T], V] \rVert_{F,\mathcal{E}} + 2\lVert [[V, T], T] \rVert_{F,\mathcal{E}} \big) ,
\end{equation}
where $d_\mathcal{E}$ is the dimension of our computational subspace of interest and $\lVert \, \cdot \, \rVert_{F,\mathcal{E}}$ is the Frobenius norm restricted to that subspace. As shown in Appendix~\ref{app:average_case_error}, we can upper bound the average-case error constant of $U_{\mathrm{tile}}$ as
\begin{equation}
    A_{\mathrm{tile}} \leq A_{\mathrm{SO}} + A_T
\end{equation}
where $A_T$ is the error contribution from the kinetic energy operator splitting (see Appendix~\ref{app:free_fermionic_operators_error} for details).

\textbf{3) Energy error} captures the difference in eigenvalues between the true Hamiltonian, $H$, and the effective Hamiltonian, $\tilde{H}$, that generates the Trotterized time evolution $U = e^{-{i\tilde{H}t}}$. Note that $\tilde{H}$ is time-step dependent -- the smaller $t$, the closer the eigenspectrum of $\tilde{H}$ to the true eigenspectrum. When performing Trotterized Hamiltonian simulation, the phase of each eigenstate will evolve according to the spectrum of $\tilde{H}$, such that this metric is especially relevant for QPE. For second-order Trotter applications, the energy error constant $C_m$ of eigenstate $m$ is defined using
\begin{equation}
    \lvert E_{m} - \Tilde{E}_m \rvert = C_m t^2 , \label{eq:absolute_phase_error}
\end{equation}
where $E_{m}$ is the true eigenenergy for state $m$ while $\Tilde{E}_m$ is the corresponding effective eigenenergy of $\tilde{H}$. Note that $C_m$ as defined here depends on $t$. We could choose to define $C_m$ in the limit of small $t$, however in practice we evaluate $C_m$ for fixed, small values of $t$ which we specify.

Calculating $C_m$ exactly is computationally expensive because it requires finding eigenvalues of $H$ and $\tilde{H}$, and constructing $\tilde{H}$ to begin with is generally infeasible. To avoid this exact construction, and to allow us to study larger systems, we introduce a computational strategy based on tensor network simulation and time series analysis. Specifically, we first use DMRG to evaluate the energy, $E_m$, and the eigenstate, $\ket{\psi_m}$, with high precision \cite{Zhai2023Block2}. To estimate the desired eigenvalues of $\tilde{H}$ we use time-dependent DMRG (TD-DMRG) algorithms to repeatedly apply the Trotterized time evolution operator, $U$, constructed as a matrix product operator, to the state $\ket{\psi_m}$. This allows us to construct the time series $g_k = \langle \psi_m | U^k | \psi_m \rangle$. From this, we may extract $\tilde{E}_m$ using a time series analysis. This time series analysis follows a similar approach to that taken in statistical phase estimation, particularly the approaches described in Refs.~\cite{Blunt2023, Wang2023}. We refer to Appendix~\ref{App:phase_error} for details. We perform DMRG and TD-DMRG using the Block2 code, using the time-dependent variational principle (TDVP) method for the latter \cite{Zhai2023Block2}. To obtain accurate results one must converge with respect to the relevant simulation parameters, particularly the DMRG bond dimension, $M$. DMRG and TD-DMRG are particularly efficient for acenes which are quasi-1D, but we also perform these computations for $n$-rhombene and $n$-triangulene with $n \leq 3$. This method allows us to significantly extend the spectral analysis of $\tilde{H}$ beyond small systems where brute force calculation is possible.

In nanographene applications as discussed in Section~\ref{sec:nanographenes}, we wish to compute the energies and character of a few low-lying eigenstates. For acenes and rhombenes, we are interested in the states $m \in \{ \mathrm{S_0}, \mathrm{S_1}, \mathrm{T_1} \}$. The low-lying eigenstates of triangulenes do not follow the same spin-state pattern and vary for different $n$, so for these, we label the ground- and first-excited state as $m \in  \{ \mathrm{gs}, \mathrm{es_1}\}$.

\textbf{4) Gap error} captures the error on energy differences between certain pairs of eigenstates. As outlined for singlet fission and inverted singlet-triplet gaps, chemistry applications are typically concerned with the relative ordering of eigenstates and the corresponding size of energy gaps. Here, we focus on energy differences between electronic eigenstates of a single molecule (and always taking a fixed geometry - that is, we study ``vertical'' excitation energies).

We denote the energy gaps between eigenstates $m$ and $n$ of $H$ as $\delta_{m,n}$, the equivalent energy gap of $\tilde{H}$ as $\tilde{\delta}_{m,n}$, and define the gap error constant, $C_{m,n}$, based on the following expression
\begin{equation}
    \lvert \delta_{m,n} - \Tilde{\delta}_{m,n} \rvert = C_{m,n} t^2 .
\end{equation}
We focus on the energy gaps between the ground state and the first few excited states. For the acenes and rhombenes, we are concerned with the error on $\delta_{\mathrm{S_0},\mathrm{T_1}}$ and $\delta_{\mathrm{S_0},\mathrm{S_1}}$, the energy gaps between the singlet ground state and either the triplet or singlet excited state respectively, while we focus on $\delta_{\mathrm{gs},\mathrm{es_1}}$ for the triangulenes. We use the numerical results on $E_m$ and $\tilde{E}_m$ (at fixed time steps, $t$) obtained from the DMRG and TD-DMRG calculations to also evaluate these gap errors (see Appendix~\ref{App:phase_error}).

\section{Trotter error results and discussion \label{sec:trotter_error_results}}
\begin{figure*}[!t]
    \centering
    \includegraphics[width=0.8\linewidth]{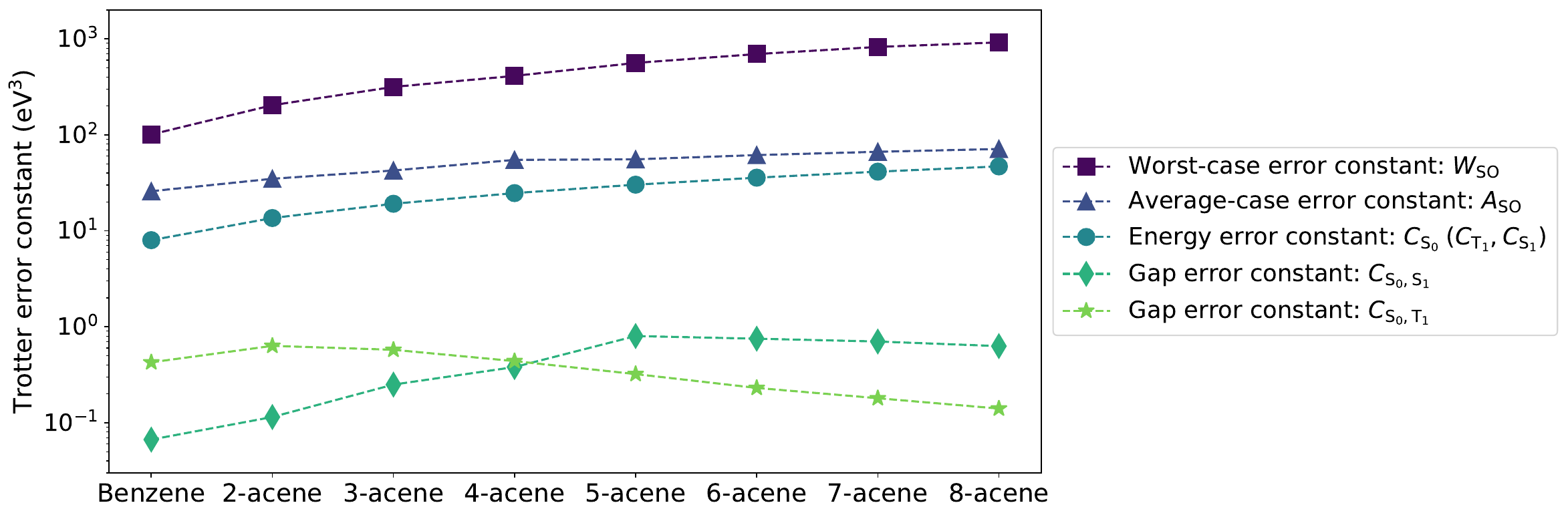}
    \caption{Comparison of 1) worst-case, 2) average-case, 3) energy and 4) gap error constants, for $U_{\mathrm{SO}}$ of $n$-acene PPP models. The Trotter error constants vary by orders of magnitude, and the gap errors are especially low in these instances. The error constants are properties of a given unitary, in this case $U_{\mathrm{SO}}$. The energy and gap errors are also eigenstate dependent, denoted by the subscripts on $C$, indicating that these errors are evaluated for the singlet ground state ($\mathrm{S_0}$) and for the gap between $\mathrm{S_0}$ and either the first-excited singlet ($\mathrm{S_1}$) or the first-excited triplet ($\mathrm{T_1}$). We write $C_{\mathrm{T_1}}$ and $C_{\mathrm{S_1}}$ in parentheses next to energy error constants to indicate that those error constants are almost equal to $C_{\mathrm{S_0}}$ (see Fig.~\ref{fig:C_estimates_acenes}). The errors should not be treated on completely equal terms and care should be taken when using them for certain algorithms and applications. The energy and gap error constants are computed at a fixed time step $t = 0.01 \; \mathrm{eV}^{-1}$.}
    \label{fig:Fig2_acene_Trotter_errors}
\end{figure*}

In this section, we present Trotter error results of $n$-acene, $n$-rhombene and $n$-triangulene PPP models. We calculate worst- and average-case errors within relevant symmetry sectors: the $N$-particle (half-filling) subspace for all systems, and $S_z = 0$ subspace for models with even $N$ and $S_z = 1/2$ subspace for models with odd $N$ (2- and 4-triangulene). The energy and gap errors are calculated using our time-series analysis for relevant low-lying eigenstates.

We show a comparison of the four error constants of $U_{\mathrm{SO}}$ for $n$-acene PPP models in Fig.~\ref{fig:Fig2_acene_Trotter_errors}. The gap error constants $C_{\mathrm{S_0},\mathrm{S_1}}$ and $C_{\mathrm{S_0},\mathrm{T_1}}$ are orders of magnitude lower than worst-case, average-case and energy errors, which highlights that in order to obtain realistic estimates of quantum simulation costs for specific applications, it is critical to consider application-specific errors. The energy-estimate errors of quantum phase estimation are proportional to the Trotter error constants, and therefore, using the same Trotter step size, we can achieve higher accuracy when estimating $\delta_{\mathrm{S_0},\mathrm{T_1}}$ and $\delta_{\mathrm{S_0},\mathrm{S_1}}$ compared to absolute energies of general states (using worst-case error) and absolute energies of low-lying eigenstates $E_{\mathrm{S_0}}$, $E_\mathrm{T_1}$ and $E_\mathrm{S_1}$. While the worst-case, average-case and energy error constants all increase gradually with system size, the gap error constants fluctuate and especially for $C_{\mathrm{S_0},\mathrm{T_1}}$ decrease significantly after $2$-acene. This means that estimating $\delta_{\mathrm{S_0},\mathrm{T_1}}$ to a given accuracy requires fewer Trotter steps for 8-acene than for 2-acene.

\begin{figure*}[t]
    \centering
    \includegraphics[width=1.0\linewidth]{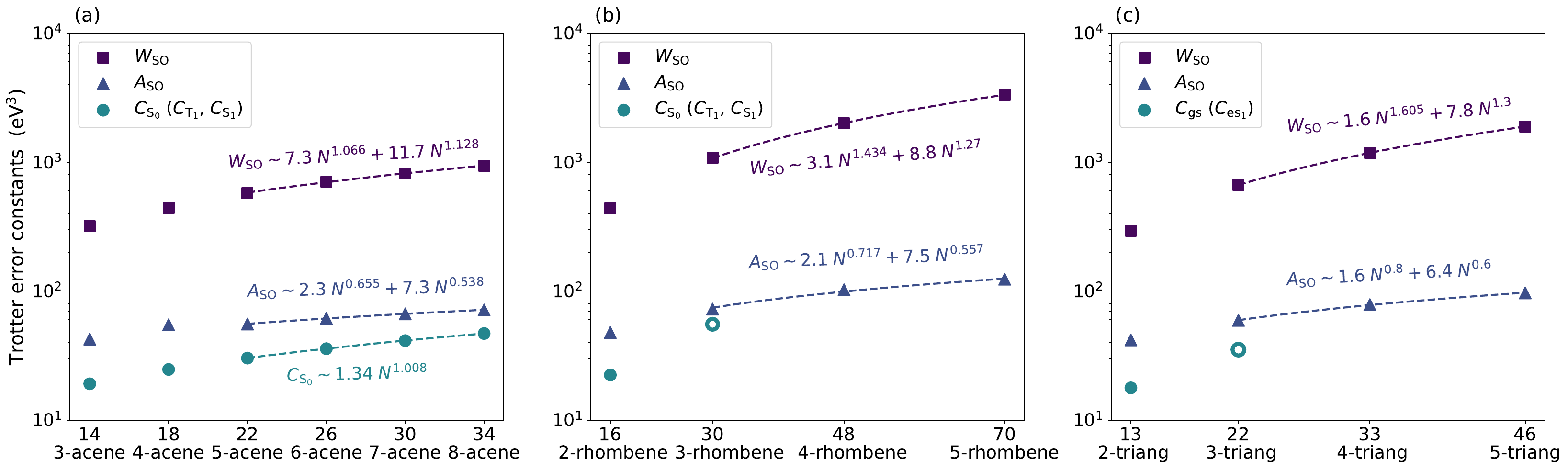}
    \caption{Comparison of worst-case ($W_{\mathrm{SO}}$), average-case ($A_{\mathrm{SO}}$) and energy ($C_m$) Trotter error constants of $U_{\mathrm{SO}}$ of (a) $n$-acene, (b) $n$-rhombene and (c) $n$-triangulene. The worst- and average-case fits ($W_{\mathrm{SO}}$ and $A_{\mathrm{SO}}$ respectively) are obtained from power-law fits of the nested commutator norms in Fig.~\ref{fig:Commutator_bounds} in Appendix~\ref{app:Trotter_errors}, whereas energy error constant fits are obtained from the data presented here. We indicate by putting $C_{\mathrm{T_1}}$, $C_{\mathrm{S_1}}$ and $C_{\mathrm{es_1}}$ in parentheses that energy error constants of excited states are similar to those of the ground states (see Figs.~\ref{fig:C_estimates_acenes}--\ref{fig:C_estimates_triangulenes}). The empty circle markers of 3-rhombene and 3-triangulene indicate that these results are not fully converged with respect to the TD-DMRG bond dimension. The energy errors are computed at fixed time step $t = 0.01 \; \mathrm{eV}^{-1}$.}
    \label{fig:Trotter_errors_acene_rhombene_triangulene}
\end{figure*}

Before investigating the gap errors in more detail, we look at the scaling of worst-case, average-case and energy errors of $U_{\mathrm{SO}}$  with system size in Fig.~\ref{fig:Trotter_errors_acene_rhombene_triangulene} for \ref{fig:Trotter_errors_acene_rhombene_triangulene}(a) $n$-acene, \ref{fig:Trotter_errors_acene_rhombene_triangulene}(b) $n$-rhombene and \ref{fig:Trotter_errors_acene_rhombene_triangulene}(c) $n$-triangulene. The worst-case errors scale as $\mathcal{O}(N^\gamma)$ with $1.06 <\gamma < 1.65$, while the average-case errors scale approximately as $\mathcal{O}(N^{\gamma/2})$ across the different systems, consistent with the quadratic improvement found in Ref.~\cite{Zhao2022}. The energy errors across the computationally feasible systems considered here are smaller than the respective average-case errors. However, the scaling of the energy error with system size for the acenes is worse than the average-case error scaling, which also seems to be the case for rhombenes and triangulenes. For these systems, it appears that the computationally inexpensive average-case error gives a good indication of the approximate magnitude of the energy errors.

\begin{figure}[t]
    \centering
    \includegraphics[width=0.75\linewidth]{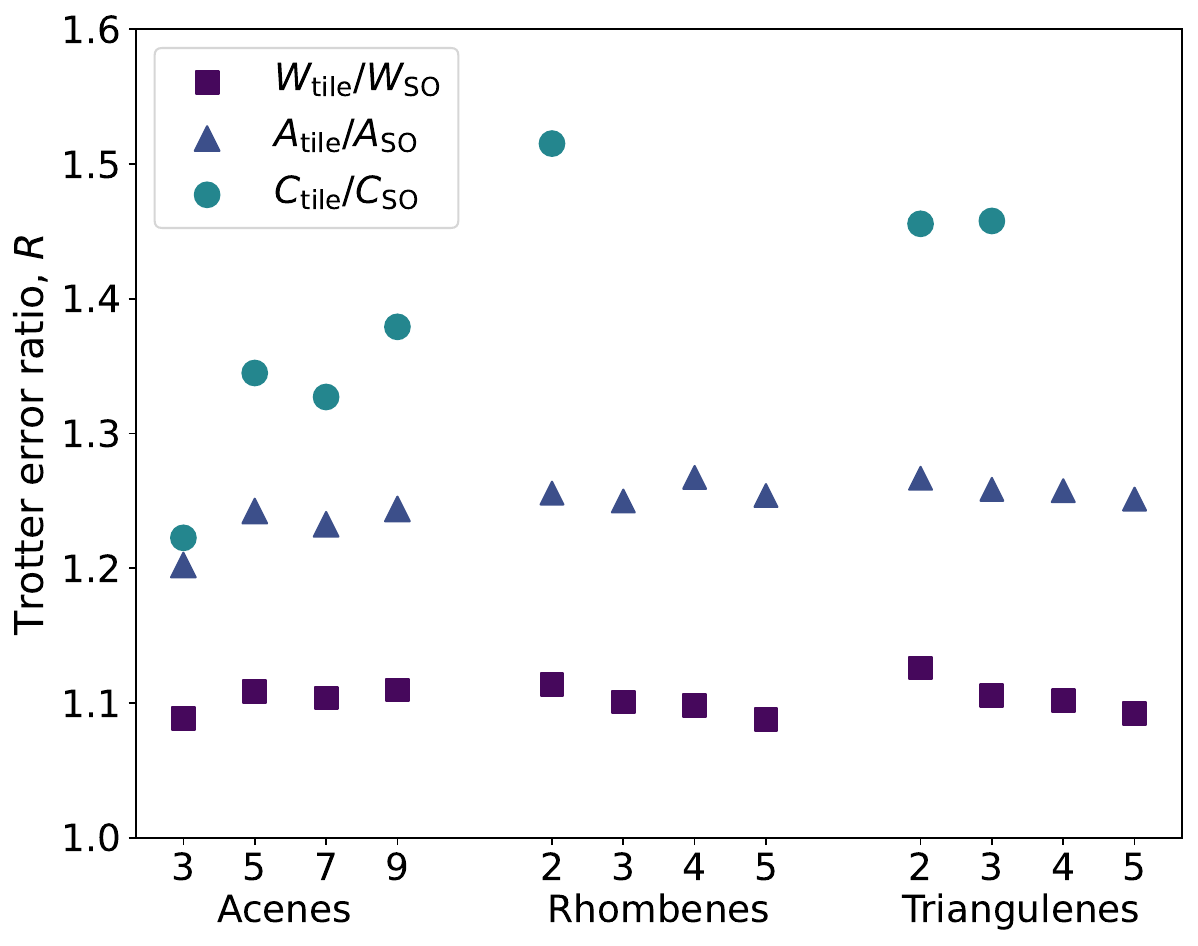}
    \caption{Worst-case, average-case and (ground-state) energy error constant ratios of $U_{\mathrm{tile}}$ over $U_{\mathrm{SO}}$ of $n$-acenes, $n$-rhombenes and $n$-triangulenes. The $U_{\mathrm{tile}}$ errors are only marginally larger, especially considering that the number of Trotter steps required for QPE and quantum dynamics using $U_{\mathrm{tile}}$ over $U_{\mathrm{SO}}$ only increases by a factor $\sqrt{R}$. The error ratio is not shown for 3-rhombene given the considerable computational resources required to calculate energy errors of $U_{\mathrm{tile}}$. We expect the 3-triangulene energy error ratio to be largely converged (with respect to the TD-DMRG bond dimension), although there may be a small convergence error.}
    \label{fig:Trotter_error_results_tile}
\end{figure}

So far, we have considered errors on $U_{\mathrm{SO}}$ which naïvely looks simpler than $U_{\mathrm{tile}}$. However, here we show that the additional Trotter error of $U_{\mathrm{tile}}$ is small, and is outweighed by its lower implementation cost compared with $U_{\mathrm{SO}}$. Fig.~\ref{fig:Trotter_error_results_tile} shows the ratio $R$ between $U_{\mathrm{tile}}$ and $U_{\mathrm{SO}}$ errors for a range of acenes, rhombenes and triangulenes. The Trotter error of $U_{\mathrm{tile}}$ is between $R \sim 1.08$ and $R \sim 1.26$ times larger than $U_{\mathrm{SO}}$ errors for the worst- and average-case across the different nanographenes. This is a negligible error contribution considering that the number of Trotter steps for (second-order) Trotterized quantum dynamics and quantum phase estimation scale as $\sqrt{W}$ or $\sqrt{A}$ using worst- or average-case errors respectively, which further suppresses the difference quadratically. The increase in the number of Trotter steps required using $U_{\mathrm{tile}}$ over $U_{\mathrm{SO}}$ for a certain application can be found simply by taking $\sqrt{R}$. The $U_{\mathrm{tile}}$ energy error is larger and varies more across the different systems, and for example simulation of 2-rhombene using $U_{\mathrm{tile}}$ requires $\sqrt{1.52}\sim1.23$ times more Trotter steps than when using $U_{\mathrm{SO}}$. However, the cost of diagonalizing general kinetic energy operators required for $U_{\mathrm{SO}}$ is also a significant contribution to the overall cost \cite{Kivlichan2020ImprovedTrotterization}. Using Givens rotations, diagonalization of general kinetic energy operators requires $\mathcal{O}(N^2)$ arbitrary rotations \cite{Kivlichan2020ImprovedTrotterization}, which leads to significantly more rotations for implementing time evolution of the kinetic energy operator in $U_{\mathrm{SO}}$ compared to $U_{\mathrm{tile}}$. We therefore believe that $U_{\mathrm{tile}}$ is both more efficient and easier to use in practice.

\begin{figure*}
    \centering
    \includegraphics[width=1.0\linewidth]{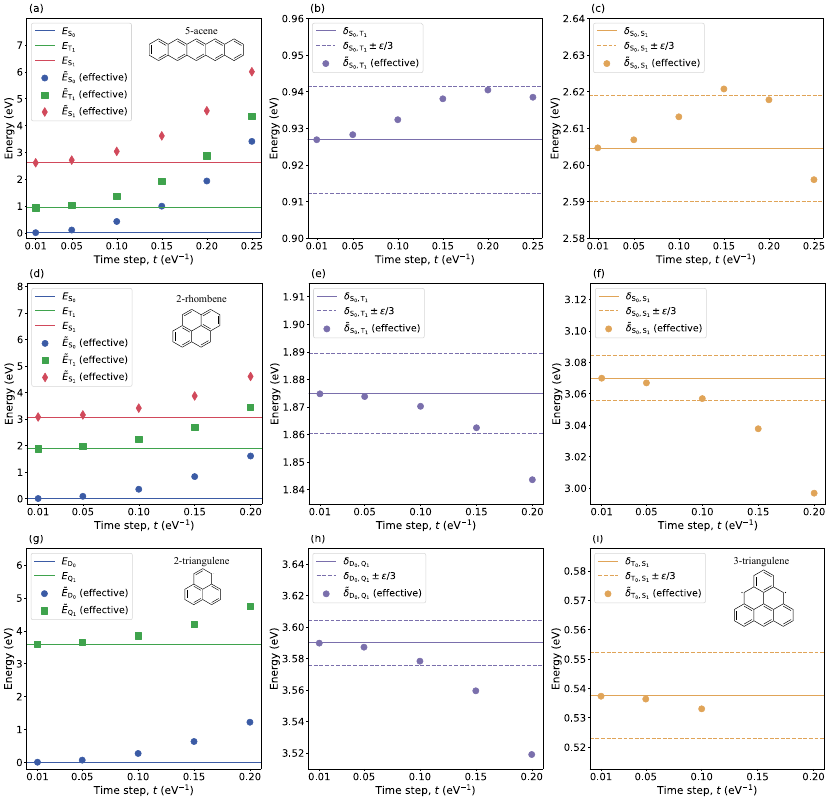}
    \caption{Exact and effective energies and energy gaps of (a), (b), (c) 5-acene and (d), (e), (f) 2-rhombene, (g), (h) 2-triangulene and (i) 3-triangulene, using the product formula $U_{\mathrm{tile}}$ as our Hamiltonian simulation approximation. Exact energies are naturally independent of the Trotter time step $t$ and remain constant (lines), while the effective energies depend on $t$ (points). All energy scales are shifted by the ground-state energy as the zero-point. (a) The solid lines show the exact energies $E$ of $\mathrm{S_0}$, $\mathrm{S_1}$ and $\mathrm{T_1}$ of the 5-acene PPP model. The points show energies, $\tilde{E}$, of the effective Hamiltonian at different values of $t$ of the same three eigenstates. At $t = 0.01 \; \mathrm{eV}^{-1}$, the effective energies are close to the exact energies (within chemical accuracy), but for larger $t$, the effective energies drift far away from the exact energies. (b) The solid line shows the exact $\mathrm{S_0}$--$\mathrm{T_1}$ energy gap, $\delta_{\mathrm{S_0},\mathrm{T_1}}$ and the dashed lines show $\delta_{\mathrm{S_0},\mathrm{T_1}} \pm \varepsilon / 3$, where $\varepsilon=1.6 \; \mathrm{mHa} \approx 0.04354 \; \mathrm{eV}$ is chemical accuracy. The points show the effective energy gap $\tilde{\delta}_{\mathrm{S_0},\mathrm{T_1}}$ for a range of $t$ where all points remain within $\varepsilon/3$ of the true energy gap. Since we allocate one-third of the total error budget to Trotter error for QPE applications \cite{Campbell_2022}, this entire $t$ range can be used for quantum simulation of the $\mathrm{S_0}$--$\mathrm{T_1}$ gap to chemical accuracy. (c) The solid line shows $\delta_{\mathrm{S_0},\mathrm{S_1}}$, the dashed lines show $\delta_{\mathrm{S_0},\mathrm{S_1}} \pm \varepsilon/3$ and the points show the effective energy gap $\tilde{\delta}_{\mathrm{S_0},\mathrm{S_1}}$ at different $t$. The trend is similar as in (b), except the effective energy gap at $t = 0.15 \; \mathrm{eV}^{-1}$ is outside the target accuracy. (d), (e) and (f) show the same quantities for the 2-rhombene PPP model, (g) and (h) show results for the 2-triangulene ground state ($\mathrm{D_0}$) and first-excited state ($\mathrm{Q_1}$), and lastly (i) shows the gap error on the ground state ($\mathrm{T_0}$) and first-excited state ($\mathrm{S_1}$) of 3-triangulene for up to $t = 0.1 \; \mathrm{eV}^{-1}$. For 3-triangulene, we used bond dimension $M=4000$ for which the effective energies are largely converged, although there may be a small remaining convergence error. Note, when performing resource estimation for phase estimation in Sec.~\ref{sec:resource_estimates} we use the trends found here and in Fig.~\ref{fig:realative_error_3_acene} (Appendix~\ref{App:phase_error}) to extrapolate to larger systems; for the larger 2D systems in particular we are unable to converge TD-DMRG simulations.} \label{fig:Trotter_relative_error_5_acene_main_text}
\end{figure*}

The small gap error constants in Fig.~\ref{fig:Fig2_acene_Trotter_errors} suggest that it is possible to reach sufficient accuracy in energy gap calculations even when using large time steps, and Fig.~\ref{fig:Trotter_relative_error_5_acene_main_text} shows results that support the claim of significant error cancellation. This figure shows the dependence of absolute energies and energy gaps of the effective Hamiltonian on the time step size of $U_{\mathrm{tile}}$ for the computationally feasible systems 5-acene, 2-rhombene, 2-triangulene and 3-triangulene, and compares them to their exact energies and exact energy gaps. We show similar plots for 3-, 7- and 9-acene in Appendix~\ref{App:phase_error}. These plots show significant Trotter error cancellation for energy gaps, which suggests that relatively large $t$ can be employed in practice. Note that this analysis requires eigenvalue estimation and TD-DMRG which quickly becomes computationally expensive for 2D systems. In the text below, we discuss the 5-acene results in detail, and similar conclusions can be drawn for the other systems.

Fig.~\ref{fig:Trotter_relative_error_5_acene_main_text}(a) shows the exact $\mathrm{S_0}$, $\mathrm{S_1}$ and $\mathrm{T_1}$ energies of the 5-acene PPP Hamiltonian (solid lines) which are naturally independent of $t$. The spectrum of the effective Hamiltonian is, however, dependent on $t$, and we show how the effective energies of the three states change as we increase $t$ (points). Not surprisingly, the effective energies move further away from the exact energies with increasing $t$. For QPE applications, we require that the effective energy differs by less than one-third of chemical accuracy ($\varepsilon=1.6 \; \mathrm{mHa} \approx 0.04354 \; \mathrm{eV}$), as we allocate approximately one-third of our total error budget to Trotter error and two-thirds to phase estimation error \cite{Campbell_2022}. Only the effective energies at $t=0.01 \; \mathrm{eV}^{-1}$ fulfill this condition, and at $t=0.05 \; \mathrm{eV}^{-1}$, the effective energy of $\mathrm{S_0}$ is in error by more than $2 \varepsilon$.

We see that the effective energies increase in a systematic manner, and lie above the true energies for all low-lying states considered in this paper. It turns out that the effective energy gaps between the states remain relatively stable across the entire $t$ range considered. Fig.~\ref{fig:Trotter_relative_error_5_acene_main_text}(b) shows the exact energy gap between $\mathrm{S_0}$ and $\mathrm{T_1}$, $\delta_{\mathrm{S_0},\mathrm{T_1}}$ (solid line) while the dashed lines show $\delta_{\mathrm{S_0},\mathrm{T_1}} \pm \varepsilon /3 $. The points show the effective $\mathrm{S_0}$--$\mathrm{T_1}$ energy gap, $\tilde{\delta}_{\mathrm{S_0},\mathrm{T_1}}$, at different $t$. The effective gap remains within the required accuracy across the entire $t$ range, which allows us to reach chemical accuracy even when using large time steps, leading to potentially significant circuit depth reductions. Fig.~\ref{fig:Trotter_relative_error_5_acene_main_text}(c) shows similar trends for the $\mathrm{S_0}$--$\mathrm{S_1}$ energy gap except at $t=0.15 \; \mathrm{eV}^{-1}$ where the energy gap does not fulfill the accuracy condition. We observe similar error-cancellation trends for other acenes as well as small 2D nanographenes in Fig.~\ref{fig:Trotter_relative_error_5_acene_main_text} and Fig.~\ref{fig:realative_error_3_acene} in Appendix~\ref{App:phase_error}.

Collecting the results from Fig.~\ref{fig:Trotter_relative_error_5_acene_main_text} as well as Fig.~\ref{fig:realative_error_3_acene} in Appendix~\ref{App:phase_error}, we estimate that using approximately $t = 0.15 \; \mathrm{eV}^{-1}$ to calculate $\delta_{\mathrm{S_0},\mathrm{T_1}}$ and $t=0.1 \; \mathrm{eV}^{-1}$ to calculate $\delta_{\mathrm{S_0},\mathrm{S_1}}$ and $\delta_{\mathrm{gs}, \mathrm{es_1}}$ (triangulenes) with QPE will allow for sufficiently accurate energy gap calculations across the systems of interest. When performing quantum phase estimation resource estimates in Section \ref{sec:resource_estimates}, we employ the more conservative time step, $t=0.1 \; \mathrm{eV}^{-1}$, across all energy gap calculations (including $\delta_{\mathrm{S_0},\mathrm{T_1}}$). Note that while we were only able to extract eigenvalue errors of the computationally efficient quasi-1D acenes as well as the small 2D systems, our results indicate that the error cancellation is consistent over a range of system sizes, and that in some cases the gap error even decreases with increasing system size. We therefore expect the observed error cancellation trends to hold for larger acenes, rhombenes and triangulenes.

\subsection{Caveats for choosing large time steps}
\label{sec:gap_errors}
The statements presented above (that we can choose large $t$ for energy gap simulations) come with three caveats.

First, choosing the optimal $t$ for computationally intractable problems will remain an educated guess. If possible, one would perform the QPE with varying time step size and monitor for convergence of results. One of the main takeaways of this paper is that using larger values of $t$ can be justified by attempting to calculate energy gaps between low-lying excited states rather than absolute energies (at least for the models considered here).

\begin{figure}[!t]
    \centering
    \includegraphics[width=0.95\linewidth]{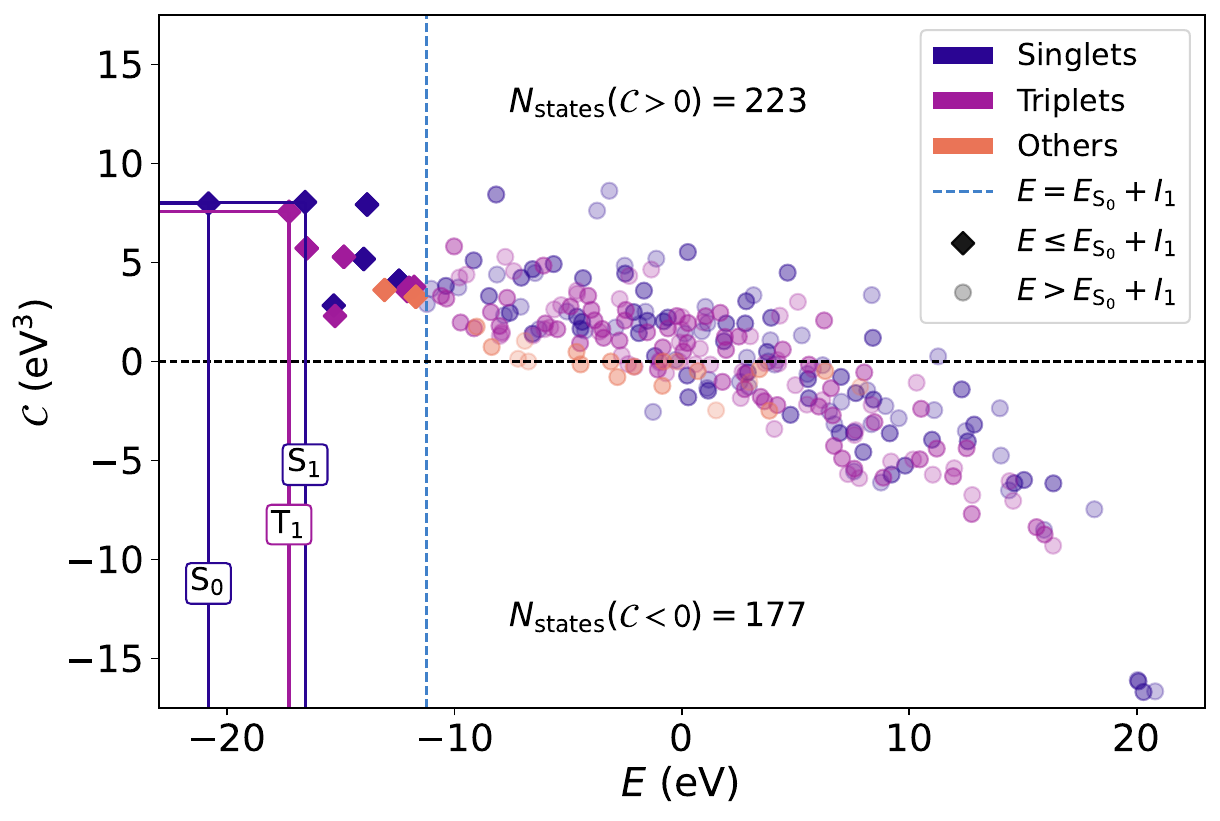}
    \caption{Plot of the magnitude- and direction-sensitive Trotter error measure $\mathcal{C}_m =(\tilde{E}_m-E_m)/t^2$ as a function of exact energies, $E_m$, of the benzene (1-acene) PPP model at half-filling using $U_{\mathrm{SO}}$ with $t=0.01 \; \mathrm{eV}^{-1}$. We note the spin character of the states, plot a horizontal line at $\mathcal{C}_m=0$ to separate the $\mathcal{C} <0$ and $\mathcal{C} > 0$ brackets and highlight states within the relevant energy range $E_{\mathrm{S_0}} \leq E < E_{\mathrm{S_0}}+I_1$ (where $I_1$ is the experimental first ionization energy of benzene \cite{GRUBB1984420}), while also plotting a vertical line at $E_{\mathrm{S_0}}+I_1$. For two different eigenstates $m$ and $n$, the gap error is larger than their respective energy errors if $\mathcal{C}_m \mathcal{C}_n <0$, while strong error cancellation (and low gap errors) happen if $\mathcal{C}_m \mathcal{C}_n > 0$ and $\mathcal{C}_m \approx \mathcal{C}_n $. Error cancellation therefore occurs between all states within the highlighted energy range, and especially $\mathrm{S_0}$, $\mathrm{S_1}$ and $\mathrm{T_1}$ have similar $\mathcal{C}$, meaning gap errors between these states are particularly low. Overall, this figure shows a clear negative correlation (Pearson correlation coefficient of $r = -0.837$ with a statistical significance probability value of $p \approx 0$ within numerical accuracy) between $E_m$ and $\mathcal{C}_m$ of the eigenstates $m$ of this system.}
    \label{fig:C_bins}
\end{figure}

Second, low gap errors are not a general characteristic for any pair of energy eigenstates, but a feature of the low-lying eigenstates investigated here. We illustrate this on the smallest nanographene system, benzene, for which we can compute both the exact eigenspectrum of $H$ and the eigenspectrum of $\tilde{H}$. For each eigenstate $m$, we determine the quantity $\mathcal{C}_m =(\tilde{E}_m-E_m)/t^2$ at $t = 0.01 \; \mathrm{eV}^{-1}$, while ensuring that we correctly compare the equivalent states of $H$ and $\tilde{H}$ respectively by sorting in terms of eigenvector overlaps. Because we take $t = 0.01 \; \mathrm{eV}^{-1}$ the Trotter error in eigenvector overlaps is very small and there is no ambiguity in determining the corresponding eigenstates, though this may be more challenging at larger time steps. The quantity $\mathcal{C}_m$ is equivalent to $C_m$ (Eq.~\eqref{eq:absolute_phase_error}) except for the absolute value, meaning $\mathcal{C}_m$ determines both the magnitude and direction of the Trotter error. Cancellation of errors only occurs between eigenstates that both lie in either the $\mathcal{C}>0$ or the $\mathcal{C}<0$ bracket, and strong error cancellation occurs between states $m$ and $n$ if $\mathcal{C}_m \approx \mathcal{C}_n$. The gap errors will be enhanced compared to energy errors for eigenstates that lie in different brackets, $\mathcal{C}_m \mathcal{C}_n < 0$. We show $\mathcal{C}_m$ for all eigenstates as a function of their exact energies, $E_m$ in Fig.~\ref{fig:C_bins}, and observe a trend that states of similar energies have similar Trotter error. We divide the points into brackets with $\mathcal{C}<0$ and $\mathcal{C}>0$ separated by a dashed line at $\mathcal{C} =0$ and highlight the spin states of the specified eigenstates (with ``Others'' including all states other than singlets and triplets). In this case, for example $\mathrm{S_0}$, $\mathrm{S_1}$ and $\mathrm{T_1}$ have strong Trotter error cancellation of energy gaps. We also highlight the states within a typical energy range of interest, $E <E_{\mathrm{S_0}}+I_1$, where $I_1$ is the first ionization energy of benzene \cite{GRUBB1984420}, showing that many chemically relevant states lie within the same region. Another way to see that cancellation cannot occur for all pairs of eigenvalues is to note that $H$ and $\tilde{H}$ differ by commutator terms, which have zero trace. Therefore, $\mathrm{Tr}(H - \tilde{H}) = 0$, meaning that $\mathcal{C}_m$ must be distributed appropriately about $0$, which we numerically confirm for our benzene example. The small gap errors between low-lying eigenstates seem to be a consequence of a correlation between Trotter errors and eigenenergies, resulting in chemically useful error cancellation between low-lying states of interest.

The third caveat is that the wrapping problem, and not the Trotter error, may set the practical upper limit on $t$ in phase estimation. This is an important caveat to keep in mind, since for quantum phase estimation, we typically require
\begin{equation}
        (E_{\mathrm{max}}-E_{\mathrm{min}}) t \le 2 \pi,
    \label{eq:wrapping_condition}
\end{equation}
to ensure that all eigenphases $E_j t$  of $e^{-iHt}$ lie within an interval of size $2\pi$, such that the eigenvalues will not ``wrap around''. For large values of $t$, the condition in Eq.~\eqref{eq:wrapping_condition} will not be satisfied. However, this condition may actually be relaxed in Trotterized phase estimation, provided that the initial state is of sufficient quality. Let us write the initial state for phase estimation in the energy eigenbasis as $| \psi \rangle = \sum_j c_j |\Psi_j \rangle $ and define a dynamic range $[ E_{\mathrm{lower}} , E_{\mathrm{upper}}]$ such that $(E_{\mathrm{upper}}- E_{\mathrm{lower}}) t = 2 \pi$.  When performing phase estimation, we shift the Hamiltonian to $H - E_c \one$ where $E_c = (E_{\mathrm{lower}}+E_{\mathrm{upper}})/2$ and seek to learn eigenvalues in the interval $[-\pi/t, \pi/t]$ for this shifted Hamiltonian. Then energies $E_j$ such that $t|E_j-E_c| \geq \pi$ are indistinguishable from energies within the dynamic range. The wrapping problem is circumvented if $t$ is sufficiently small that the following holds
\begin{align} \label{eq:wrap}
  | c_m |^2 & \gg \sum_{j:   t | E_j - E_c | \geq \pi }|c_j|^2
\end{align}
for all $m$ states for which we wish to learn the energy.  Note that $E_{\mathrm{lower}}$ and $E_{\mathrm{upper}}$ can be defined implicitly by choosing a shift $E_c$ for the Hamiltonian and a time step, $t$. The higher $t$ is, the lower the resource overhead of phase estimation; whereas the smaller $t$ is, the more confident we can be that Eq.~\eqref{eq:wrap} is satisfied. We should also choose $E_c$ such that a lower bound estimate of $E_0$ is within the dynamic range. In practice, Eq.~\eqref{eq:wrap} can be satisfied even for relatively large $t$ by choosing a good initial state. For example, consider the case where $| \psi \rangle$ has been chosen variationally to target the ground state, such that
\begin{equation}
    \langle \psi | H | \psi \rangle = \sum_j |c_j|^2 E_j
\end{equation}
is close to the ground-state energy. This then implies that the spectral weight on high-energy states is close to zero. In this case, the wrapping of high-energy states will not prevent accurate phase estimation. This is often assumed but not discussed, for example in previous papers on phase estimation for the Hubbard model with extensive target accuracies \cite{Kivlichan2020ImprovedTrotterization, Campbell_2022, Bay-Smidt2025}, where Eq.~\eqref{eq:wrapping_condition} will also be violated. However, one should keep in mind that increasingly large values of $t$ imply an increasingly strict requirement on initial state preparation.

\section{Quantum phase estimation resource estimates \label{sec:resource_estimates}}

\begin{figure*}[t]
    \centering
    \includegraphics[width=1.0\linewidth]{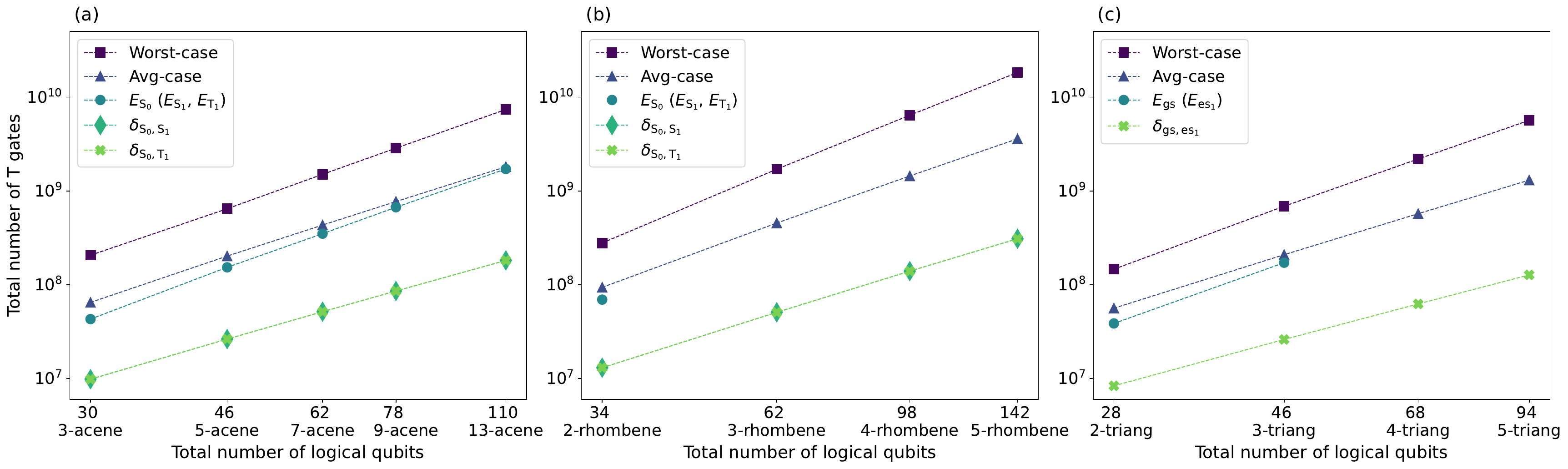}
    \caption{T gate and logical qubit resource estimates of (a) $n$-acene, (b) $n$-rhombene and (c) $n$-triangulene PPP model quantum phase estimation simulations, using different Trotter error measures and with $U_{\mathrm{tile}}$ as our Hamiltonian simulation operator. This quantum phase estimation scheme requires one phase qubit in addition to the $2N$ system qubits for nanographene PPP models with $N$ carbon atoms, and a qubit for repeat-until-success synthesis of arbitrary rotations \cite{Bocharov2015EfficientCircuits}. We use the Trotter error constants for the worst-case, average-case and energy error to perform the T gate costings, see Eqs.~\eqref{eq:N_PE_worst_avg_absolute}--\eqref{eq:N_T_total_worst_avg_absolute} in Appendix~\ref{app:phase_estimation_costing}. We estimated the T gate cost of 13-acene based on the fit of $C_{\mathrm{tile}}$ presented in Fig.~\ref{fig:C_estimates_acenes}, while for the rhombenes and triangulenes, we do not have enough data for the energy error constant to confidently extrapolate to larger systems. For all energy gap estimates, we fix $t=0.1 \; \mathrm{eV}^{-1}$ based on the eigenvalue analysis presented in Fig.~\ref{fig:Trotter_relative_error_5_acene_main_text}, which results in the same cost of estimating $\delta_{\mathrm{S_0},\mathrm{T_1}}$ and $\delta_{\mathrm{S_0},\mathrm{S_1}}$ in (a) and (b). We use Eq.~\eqref{eq:N_PE_relative} and Eq.~\eqref{eq:N_T_step_relative} to obtain the total T gate costs after fixing $t$. Estimating the energy gap $\delta_{m,n}$ requires two independent rounds of quantum phase estimation using the same $t$, where the first round determines the effective energy $\tilde{E}_m$ and the second round retrieves $\tilde{E}_n$. Note that performing eigenvalue error estimation of the large 2D systems ($n$-rhombene with $n\geq3$ and $n$-triangulene with $n\geq4$) with the classical computational method applied here would be very challenging. The T gate estimates for calculating energy gaps of these large 2D systems are therefore based on extrapolation of the error cancellation trends found for the acenes and the small rhombenes and triangulenes. The T gate counts in this figure do not use Hamming weight phasing.}
    \label{fig:QPE_resource_estimates}
\end{figure*}

Computational quantum chemistry is often concerned with determining the energy and character of given eigenstates, and especially energy differences between such states are relevant quantities for a large range of applications, including singlet fission, singlet-triplet inversion and magnetic properties of nanographenes as discussed in Section~\ref{sec:nanographenes}. Here, we focus on quantum simulation for determining energies and energy gaps between electronic states with certain spin properties of molecules with fixed geometries.

Quantum phase estimation (QPE) is an algorithm to estimate eigenvalues of a given Hamiltonian provided that an initial state with sufficient overlap with the target state can be prepared. The state preparation problem is not discussed here, and all resource estimates disregard the cost of state preparation. However, we note that preparing states within certain symmetry sectors of the Hamiltonian that are conserved by each Trotter fragment, for example $N$-particle and spin ($S_z$ or $S^2)$, may be particularly useful for targeting low-lying eigenstates and gaps between states within specific symmetry sectors, for example the gap between low-lying singlets and triplets.

We assume an adaptive phase estimation scheme \cite{Higgins_2007, Berry2009, Kivlichan2020ImprovedTrotterization} that requires a single ancilla qubit in addition to the $2N$ system qubits of nanographene PPP models. We use the second-order Trotter formula $U_{\mathrm{tile}}$ as our Hamiltonian simulation operator, for which we can implement directionally-controlled evolution \cite{Berry2009, Blunt2024}, which does not require additional qubits or non-Clifford operations for $U_{\mathrm{tile}}$. For synthesis of arbitrary rotations, we use the repeat-until-success scheme which requires an additional qubit \cite{Bocharov2015EfficientCircuits}. Using this adaptive phase estimation scheme, estimating an eigenvalue to accuracy $\varepsilon_{\mathrm{PE}}$ requires time evolution for a total time of \cite{Kivlichan2020ImprovedTrotterization}
\begin{equation}
    t_{\mathrm{tot}} = \frac{0.76\pi}{\varepsilon_{\mathrm{PE}}},
    \label{eq:t_total}
\end{equation}
which requires us to apply $U_{\mathrm{tile}}$ $N_{\mathrm{steps}} = t_{\mathrm{tot}}/t$ times. We give the formulas for $N_{\mathrm{steps}}$, including how the total error budget $\varepsilon$ is divided between phase estimation, Trotter and rotation synthesis errors, in Appendix~\ref{app:phase_estimation_costing}.

In practice, we perform the resource estimates in two ways, which differ only in how $t$ is fixed: 1) using worst-case, average-case and energy errors, we calculate the Trotter error constant and use this to directly estimate the required number of Trotter steps (indirectly fixing the time step), and 2) using gap errors, we fix $t=0.1 \; \mathrm{eV}^{-1}$ based on the conditions discussed in Fig.~\ref{fig:Trotter_relative_error_5_acene_main_text} and then determine the number of required Trotter steps. Our resource estimates include all non-Clifford gates.

\begin{figure}[t]
    \centering
    \includegraphics[width=0.8\linewidth]{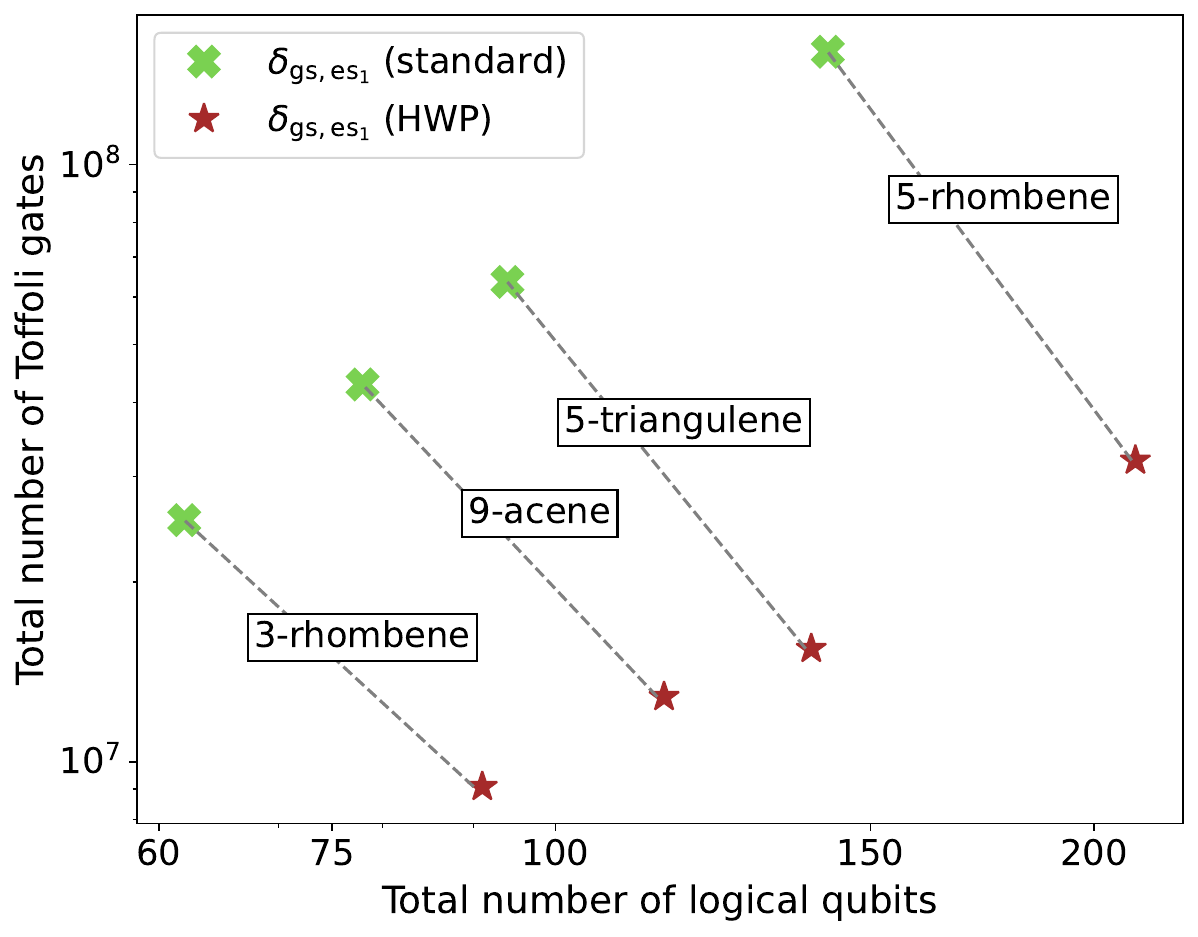}
    \caption{Toffoli gate and logical qubit resource estimates for calculating the energy gap between the ground-state (gs) and the first-excited state ($\mathrm{es_1}$) of selected systems with two different implementations of QPE: standard implementation of $U_{\mathrm{tile}}$  (as in Fig.~\ref{fig:QPE_resource_estimates}) and an implementation using HWP \cite{Gidney2018halvingcostof, Campbell_2022} to reduce the total non-Clifford cost at the cost of additional ancilla qubits.}
    \label{fig:QPE_resource_estimates_HWP}
\end{figure}

Fig.~\ref{fig:QPE_resource_estimates} shows logical qubit counts and T gate estimates for QPE to chemical accuracy of \ref{fig:QPE_resource_estimates}(a) $n$-acene, \ref{fig:QPE_resource_estimates}(b) $n$-rhombene, and \ref{fig:QPE_resource_estimates}(c) $n$-triangulene PPP models, considering the different Trotter errors. The non-Clifford cost at a fixed $N$ is comparable across the three systems, making quantum simulations of nanographenes relatively unaffected by their topology. These resource estimates suggest that evaluating energy gaps between low-lying eigenstates reduces the T gate cost by approximately an order of magnitude compared to absolute energy estimation of low-lying eigenstates. Note that estimating the energy gap between states $m$ and $n$, $\delta_{m,n}$, requires two independent rounds of quantum phase estimation using the same time step size, where the first round estimates the effective energy of state $m$ and the second round estimates the effective energy of state $n$. Fig.~\ref{fig:QPE_resource_estimates} also shows that worst-case errors overestimate resource requirements by up to two orders of magnitude for the systems and applications considered here, which confirms the importance of application-specific Trotter error analysis for more realistic Trotterized quantum simulation resource estimates.

There are certain algorithmic choices and trade-offs that can be made to further reduce gate counts. We illustrate this by comparing the standard QPE implementation in Fig.~\ref{fig:QPE_resource_estimates} to a QPE implementation where we use Hamming weight phasing (HWP) to reduce the total number of arbitrary rotations at the cost of additional ancilla qubits. In Fig.~\ref{fig:QPE_resource_estimates_HWP}, we compare the Toffoli resource requirements of QPE that uses a standard implementation of $U_{\mathrm{tile}}$, to QPE with HWP (see Appendix~\ref{sec:QPE_with_HWP}). We focus on the four nanographenes indicated, and the dashed lines connect same-system costings with either standard- or HWP-QPE. We estimate that the energy gaps between the ground state and low-lying excited states of relatively large 2D systems such as 5-triangulene and 5-rhombene (PPP models) can be computed to chemical accuracy by performing two independent QPE runs with less than $3.2 \times 10^7$ Toffoli gates  each (disregarding the cost of state preparation). We note, however, that while non-Clifford gate counts are the appropriate metric in the Pauli-based computation model \cite{Bravyi2016, Litinski2019}, in other models it may be more appropriate to assess alternative metrics such as quantum volume. While HWP decreases the non-Clifford gate count, it may increase the depth and volume of the circuit due to the additional ancillas and reduced parallelizability. This was recently studied within the FLASQ cost model \cite{Huggins2025}, demonstrating these potential issues. For this reason, the non-HWP approach may be preferable within many models of computation, particularly for early fault tolerance.

Note that our TD-DMRG calculations in Figs.~\ref{fig:C_estimates_rhombenes} and \ref{fig:C_estimates_triangulenes} in Appendix~\ref{App:phase_error} show that even relatively small 2D systems (3-triangulene and 3-rhombene) require large bond dimensions ($>3800$) for convergence, suggesting that larger rhombenes and triangulenes are non-trivial to simulate classically, even using optimized DMRG and TD-DMRG codes \cite{Zhai2023Block2}, although we do not perform a benchmarking of alternative classical methods such as PEPS here \cite{VerstraeteCirac2004, Alkabetz2021}.

\section{Conclusion}
In this paper, we have proposed quantum simulation of nanographene $\pi$-systems as a set of chemically-relevant and scalable problems that span the gap between early fault-tolerant and large-scale fault-tolerant quantum computing applications. We have presented a detailed analysis of Trotter errors and shown that different measures of Trotter error vary by orders of magnitude, such that choosing the error measure appropriate to the application can significantly reduce resource requirements. As part of this analysis, we studied the Trotter error on vertical excitation energies, and showed that the computation of such energy gaps appears to be a particularly suitable quantum simulation application. In particular, we showed numerically that there is a very significant degree of Trotter error cancellation in energy estimates for low-lying states, resulting in an order of magnitude circuit depth reduction for estimating energy gaps rather than absolute energies. This result is significant because chemistry applications typically rely on energy differences and not absolute energies. This analysis was achieved by a numerical approach based on tensor network simulation and time series analysis, allowing us to perform accurate spectral analysis of product formulas for systems beyond the reach of exact diagonalization.

We estimate that quantum simulation of energy gaps between the ground- and excited-states of the Pariser--Parr--Pople model of large 2D nanographenes (up to 140 spin orbitals) requires circuits with $< 3.2 \times 10^7$ Toffoli gates. This is significantly lower than many current estimates for Trotterized quantum phase estimation of \emph{ab initio} chemistry problems in the literature. We note that performing QPE for the Hubbard model has been estimated to require $\sim 10^6$ non-Clifford gates (dependent on the target accuracy) \cite{Campbell_2022, Kan2025, Bay-Smidt2025}, and based on this low non-Clifford count, the Hubbard model has been widely considered a good application for early fault-tolerant quantum computers \cite{Campbell_2022, Kan2025, Bay-Smidt2025, Toshi2025, Akahoshi2025, Chung2026}. Our results show that low-lying energy gaps for non-trivial 2D nanographenes within the PPP model can be obtained within chemical accuracy with only slightly greater non-Clifford counts. This suggests that nanographenes are a suitable next application to consider, bridging the gap between earlier and later quantum simulation on fault-tolerant quantum computers. Despite being a semi-empirical model, the PPP model can be used to predict the optical and dynamical properties of a range of nanographene molecules of scientific and industrial importance \cite{Chakraborty_2013, Lambie2025, sony_correlated_2005, Sony2007, chiappe_can_2015, bhattacharyya_pariserparrpople_2020, Bostrom2018}.

We expect that the resource estimates presented in this paper can be reduced in a number of ways. For example, more efficient Clifford+T rotation synthesis can be performed using mixed approximations \cite{Campbell2016, Hastings2017, Kliunchnikov2023}, while lower-depth phase estimation circuits can be achieved using statistical phase estimation techniques \cite{Lin2022, Wan2022, Dutkiewicz2022, Wang2023, Blunt2023, Ding2023}, typically at the expense of a higher total runtime. Higher-order product formulas, such as fourth-order formulas obtained by Suzuki's recursive construction \cite{Suzuki1991GeneralPhysics} or the optimized eighth-order formulas of Ref.~\cite{Morales2025}, may reduce the cost of absolute energy estimates. However, it remains open whether higher-order formulas show a similar degree of error cancellation.

In our analysis, we noted three relevant caveats for choosing large Trotter step sizes. We briefly highlight these caveats and note where further work is required to understand the generality and significance of the results presented in this paper. First, for computationally intractable problems, it is not possible to know the optimal Trotter step size. This paper provides evidence that using larger step sizes can be justified when calculating energy differences between low-lying eigenstates rather than absolute energies, but further work is required to understand the generality and limitations, especially for other Hamiltonians and Trotter schemes than the ones considered here.  Second, the observed Trotter error cancellation is not general for any pair of eigenstates, but appears to be a feature for pairs of eigenstates that are close in energy. For benzene we observed a strong negative correlation between the eigenenergy and the Trotter error of a given eigenstate, but identifying additional patterns in eigenstates with strong error cancellation is an interesting direction for future work, which would allow us to more reliably extend our findings to classically intractable problems. Third, the wrapping problem of quantum phase estimation may set the upper limit on the Trotter step size in practice, or otherwise set stricter requirements on the quality of the initial state.

In this paper, we focused on vertical excitation energies between relevant low-lying electronic states of individual molecules with fixed nuclear coordinates. However, this does not cover all relevant applications where energy differences are the primary quantities of interest. Energy differences within the same electronic state but at different nuclear geometries are also extremely important, as they define the potential energy landscape that underlies the description of many chemical phenomena. We expect that there may also be Trotter error cancellation in this scenario, but this is left for future work.

In conclusion, this paper shows that considering details of chemically-relevant applications and exploiting error cancellation can lead to substantial reductions in resource requirements. Similarly to how approximate computational methods in classical computing have proven successful for a broad range of applications, we expect the practical utility of quantum computers for simulating molecules and materials to benefit significantly from exploiting error cancellation in the approximations introduced in quantum simulation.

\section*{Data availability}
Data files for the TD-DMRG calculations and code to generate and obtain results for the kinetic energy operator sections of $U_{\mathrm{tile}}$ are uploaded to Zenodo~\cite{zenodo_data} with the DOI identifier \url{https://doi.org/10.5281/zenodo.19915016}. \vspace{0.2cm}

\section*{Acknowledgements}
This work is supported by the Novo Nordisk Foundation, Grant number NNF22SA0081175, NNF Quantum Computing Programme. \vspace{0.2cm}

\section*{Author contributions}
A.J.B.S. conceived the main idea of exploiting Trotter error cancellation, performed most of the calculations and analysis, and wrote the majority of the manuscript. N.G. provided guidance and supervision, in particular on the DMRG and TD-DMRG calculations. M.D.F. contributed to the description of nanographenes and the details of the PPP model. E.T.C. contributed to the mathematical analysis, in particular of the free fermionic operator errors, and provided general feedback. N.S.B. contributed substantially to the development of the numerical Trotter error analysis based on DMRG and TD-DMRG, ran QMC simulations to bound the worst-case Trotter error, and helped supervise the project. G.C.S. supervised the project and provided detailed feedback on the final manuscript. N.G. and G.C.S. acquired funding and computational resources. All authors contributed to writing and reviewing the manuscript.

After acceptance for publication in \emph{Quantum}, the authors used Claude Opus 5.5 (Anthropic) and GPT-6.2 Astra (OpenAI) to proofread the final version and check it for errors and inconsistencies. All changes were reviewed by the authors, who take full responsibility for the content.

\bibliographystyle{quantum} \bibliography{main_final}

\onecolumn
\appendix
\section{Symmetry shift of potential energy operator \label{app:sym_shift}}
The potential energy operator of the Pariser--Parr--Pople model is defined as
\begin{equation}
     V = u \sum_i^N n_{i\uparrow}n_{i\downarrow} + \sum_{i<j} v_{ij} (n_{i} - \one) (n_{j} - \one),
\end{equation}
where $u$ is the on-site interaction parameter and $v_{ij}$ is the interaction parameter between electrons on sites $i$ and $j$ at distance $r_{ij}$, here given by the Ohno potential
\begin{equation}
    v_{ij} = \frac{u}{{\sqrt{1+\alpha r_{ij}^2}}},
\end{equation}
where $\alpha$ is a fixed parameter. We consider PPP models of nanographenes with $N$ carbon atoms and $2N$ spin orbitals. Typical pre-processing methods for modern Hamiltonian simulation techniques require minimization of the 1-norm \cite{Loaiza2023}, but for the implementation cost of Trotterization as performed in this paper, this is less relevant. Instead, to improve Trotterized quantum simulation, our goal is to remove as many terms as possible by shifting $V$ by operators proportional to the number operator, $\hat{N}=\sum_i^N \sum_\sigma n_{i\sigma}$, and the number operator squared, employing similar symmetry shifts as the ones presented in Ref.~\cite{Loaiza2023}. We define our shifted potential energy operator as
\begin{equation}
    V' = V + c_1 \hat{N} + c_2 \hat{N}^2,
\end{equation}
and choose $c_1$ and $c_2$ to minimize the number of terms. We choose these parameters in two separate steps; first we shift by $c_2 \hat{N}^2$, and then we shift the resulting operator by $c_1 \hat{N}$.

First, let $v_{\mathrm{freq}}$ be the most frequently occurring parameter, $v_{ij}$, in $V$. Choosing $c_2$ as
\begin{equation}
    c_2 = - \frac{1}{2}v_{\mathrm{freq}},
\end{equation}
allows us to remove all terms of the form $v_{\mathrm{freq}} n_{i\sigma} n_{j\sigma'}$, and shifts the rest of the terms.

Second, define $V_{\mathrm{Z}}$ as all single ${Z}$ terms in the JW-transformed $(V+c_2\hat{N}^2)$. Then, write this operator as $V_Z=\sum_i^N h_i Z_i$, let $h_{\mathrm{freq}}$ be the most frequently occurring $h_{i}$ coefficient, and choose $c_1$ as
\begin{equation}
    c_1 = 2 h_{\mathrm{freq}},
\end{equation}
removing all terms of the form $h_{\mathrm{freq}} Z$ from the JW-transformed potential energy operator. In fact, due to the structure of the potential energy operator in the PPP model, all single-$Z$ terms have the same coefficient, and we can therefore remove them all using this shift.

These symmetry shifts will lead to overall reductions in the number of terms in the JW-transformed potential energy term, and the effects of these shifts are summarized in Table~\ref{tab:number_of_terms_V_and_V_shifted} in the main text. We use this symmetry-shifted potential energy term when performing resource estimates of single Trotter steps, see Table~\ref{tab:per_trotter_step_costs} in Appendix~\ref{app:phase_estimation_costing}.

\section{Trotter errors of SO- and tile-Trotter decompositions \label{app:Trotter_errors}}
We consider the following two Trotter decompositions of PPP Hamiltonians
\begin{align}
     U_{\mathrm{SO}} &= e^{-iVt/2} e^{-iTt} e^{-iVt/2}, \label{eq:U_SO_app} \\
     U_{\mathrm{tile}} &= e^{-iVt/2} \prod_{s=1}^S e^{-iT_s t/2} \prod_{s=S}^1 e^{-iT_s t/2}  e^{-iVt/2}, \label{eq:U_SO_tile_app_v1}
\end{align}
where the subscript $\mathrm{SO}$ denotes the split-operator decomposition and tile refers to the tile Trotterization scheme of Ref.~\cite{Bay-Smidt2025}, which is a generalization of plaquette Trotterization \cite{Campbell_2022}. In the split-operator approach, we perform time evolution of the potential energy ($V$) and kinetic energy ($T$) operators separately, assuming that each of these can be implemented without additional error. This is trivially true for $V$ (all terms commute), while it requires diagonalization of the kinetic energy operator between each application of $e^{-iTt}$. This transformation will not be described in detail here, but we note that it can be achieved in the general case using networks of Givens rotations, which in general is expensive \cite{Kivlichan2020ImprovedTrotterization}. In the plaquette \cite{Campbell_2022} and tile Trotterization \cite{Bay-Smidt2025} approach, the kinetic energy operator is further split into $S$ sections that each can be diagonalized individually with relatively low non-Clifford gate count at the cost of an additional Trotter error contribution compared to the SO scheme. We refer to Ref.~\cite{Bay-Smidt2025} for a detailed description of the tile Trotterization scheme. We label the Trotter approximation of $e^{-iTt}$ as $U_T$ and define it for the second-order tile Trotterization step as
\begin{equation}
    U_T = \prod_{s=1}^S e^{-iT_s t/2} \prod_{s=S}^1 e^{-iT_s t/2}, \label{eq:U_T}
\end{equation}
where $T = \sum_s^S T_s$. This allows us to write $U_{\mathrm{tile}}$ as
\begin{equation}
    U_{\mathrm{tile}} = e^{-iVt/2} U_T e^{-iVt/2}. \label{eq:U_SO_tile_app_v2}
\end{equation}

In the following sections, we show how to evaluate the worst-case, average-case, energy and gap errors for both $U_{\mathrm{SO}}$ and $U_{\mathrm{tile}}$. Note that the Trotter errors are invariant to the symmetry shifts performed in Appendix~\ref{app:sym_shift} when considering a fixed number of electrons, and any other diagonal symmetry shifts \cite{MartinezMartinez2023assessmentofvarious}. In practice, we perform the worst- and average-case error calculations using the original Hamiltonian, while we add a constant energy shift to $H$ for the calculations of energy and gap errors to symmetrize the spectrum around zero. This symmetrization simplifies certain aspects of the TD-DMRG and time-series analysis related to the wrapping problem, which is discussed further in Appendix~\ref{App:phase_error}. Note this symmetrization is not essential for the calculations performed here, but utilizing constant energy shifts to target certain energy intervals of a Hamiltonian may be useful in general. For more details on this, see the discussion on the wrapping problem in Section~\ref{sec:gap_errors}.

\subsection{Worst-case error \label{app:worst_case_error}}

\begin{figure*}[t]
    \centering
    \includegraphics[width=1.0\linewidth]{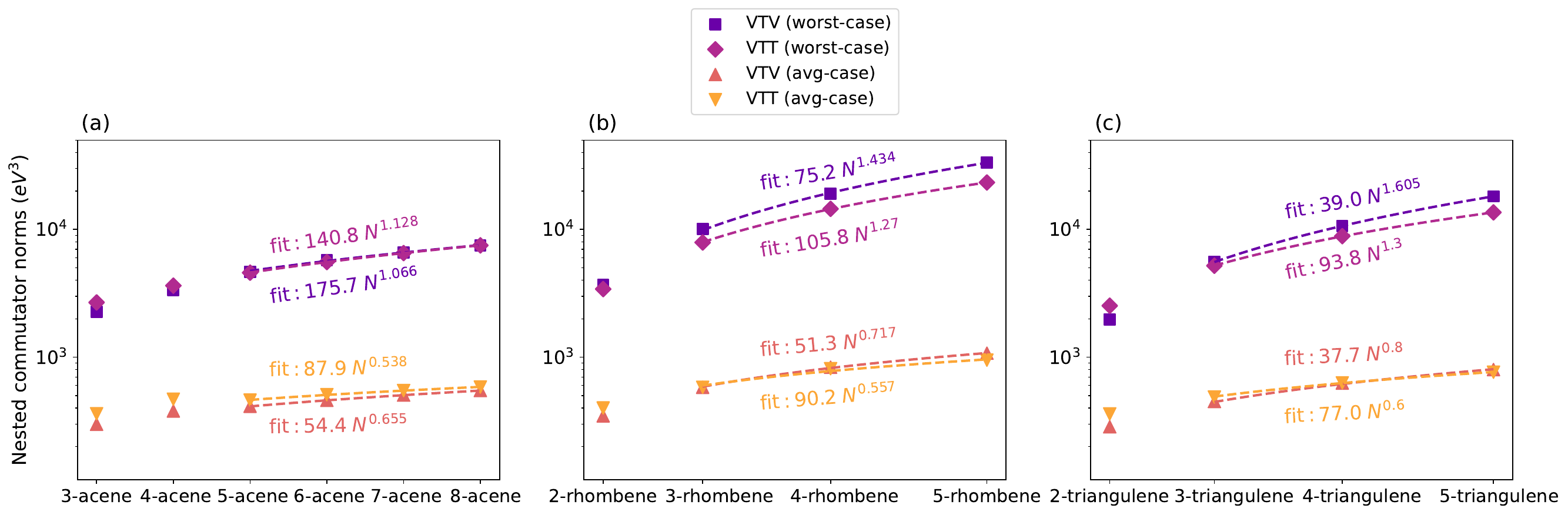}
    \caption{Spectral norms (worst-case) and normalized Frobenius norms (avg-case) of the nested commutators $[[V,T],V]$ and $[[V,T],T]$ of (a) $n$-acene, (b) $n$-rhombene and (c) $n$-triangulene PPP models. The worst-case norms are calculated using a Monte Carlo approach \cite{blunt2025montecarloapproachbound}, and the average-case norms are evaluated by sampling as shown in Eq.~\eqref{eq:norm_frob_norm_sampling}. We fit a power-law to the large systems (avoiding edge-effects), and plot the fit along with the nested commutator norms. The data and the respective standard errors of each point in these plots are shown in Table~\ref{tab:nestes_commutator_norms}.}
    \label{fig:Commutator_bounds}
\end{figure*}

\begin{table*}[t]
\resizebox{\textwidth}{!}
{\begin{tabular}{lrrrrrrrrrrrrrrrr}
\hline
\hline
& \multicolumn{6}{c}{$n$-acene} & & \multicolumn{4}{c}{$n$-rhombene} & & \multicolumn{4}{c}{$n$-triangulene} \\
\cline{2-7}
\cline{9-12}
\cline{14-17}
& 3 & 4 & 5 & 6 & 7 & 8 & $\;$ & 2 & 3 & 4 & 5 & $\;$ & 2* & 3 & 4* & 5  \\
\hline
\noalign{\vskip 1pt}
$\lVert \OVTV \rVert$ &  2665.0 & 3338.0 & 4655.1 &  5761.3 & 6600.4 &  7494.4 & &  3694.6 & 10115.9 & 19073.2 &  33342.7 & & 1976.0 & 5582.4 & 10608.0 &  18145.0 \\
SE & 0.11 & 0.10 & 0.20 & 0.26 & 0.14 & 0.49 & & 0.29 & 0.88 & 2.3 & 30 & & 0.17 & 0.36 & 0.78 & 1.35 \\
\hline
\noalign{\vskip 1pt}
$\lVert \OVTT \rVert$ & 2684.0 & 3631.1 & 4592.5 & 5557.8 & 6531.1 & 7508.7 & & 3415.5 & 7907.7 & 14510.8 & 23317.9 & & 2528.4 & 5203.5 & 8864.4 & 13612.9 \\
SE & 0.87 & 0.47 & 0.86 & 0.62 & 0.59 & 0.51 & & 0.22 & 1.3 & 3.5 & 3.4 & & 0.17 & 0.69 & 1.2 & 2.8 \\
\hline
\noalign{\vskip 1pt}
$ \frac{\lVert \OVTV \rVert_F}{\sqrt{d}} $ & 298.6 & 379.0 & 412.7 & 459.6 & 505.5 & 548.6 & & 347.2 & 581.0 & 835.3 & 1075.1 & & 285.3 & 449.9  & 627.4 & 800.2 \\
SE & 0.28 & 0.40 & 0.36 & 0.36 & 0.39 & 0.38 & & 0.33 & 0.45 & 0.70 & 0.66 & & 0.31 & 0.42 & 0.57 & 0.54 \\
\hline
\noalign{\vskip 1pt}
$\frac{\lVert \OVTT \rVert_F}{\sqrt{d}}$ & 361.1 & 469.6 & 461.2 & 509.1 & 548.1 & 583.9 & & 400.6 & 581.5  & 813.4 & 948.3 & & 359.7 & 487.5 & 630.1 & 763.3 \\
SE & 0.28 & 0.43 & 0.29 & 0.34 & 0.37 & 0.35 & & 0.25 & 0.37 & 0.62 & 0.55 & & 0.27 & 0.28 & 0.42 & 0.48 \\
\hline
\hline
\end{tabular}}
\caption{Nested commutator norms for worst- and average-case Trotter error calculations of $n$-acene, $n$-rhombene and $n$-triangulene PPP models. We specify the relevant norms of $\OVTV$ and $\OVTT$ along with the standard error (SE) of each norm. The unit on all these quantities is $\mathrm{eV}^3$. The spectral norms obtained from the Monte Carlo sampling are upper bounds to the true norm, while the normalized Frobenius norms are obtained by sampling. We obtain a standard error of less than 1$\%$ for all these calculations, which is more than sufficient when using these norms for resource estimates. We note that the 4-acene average-case $\OVTT$ breaks the otherwise clear trend of increasing norms with system size, but after rerunning our sampling, we believe this is an artifact of this system. These calculations have been performed at half-filling ($N$ electrons in $2N$ spin orbitals) and in the $S_z = 0$ subspace for all systems except for 2- and 4-triangulene as highlighted with * where calculations were performed in the $S_z = 1/2$ subspace.}
\label{tab:nestes_commutator_norms}
\end{table*}

The worst-case Trotter error of the exact Hamiltonian simulation unitary, $\mathcal{U}=e^{-iHt}$, and its Trotterized approximation, $U$, is given by
\begin{equation}
    \mathcal{W}(\mathcal{U},U) =\lVert \mathcal{U}-U \rVert,
\end{equation}
where $\lVert \, \cdot \, \rVert$ is the operator norm (or spectral norm). For the second-order split-operator (SO) Trotter approximation, $U_{\mathrm{SO}}$, the worst-case Trotter error is given by
\begin{equation}
    \lVert  \mathcal{U} - U_{\mathrm{SO}} \rVert \leq W_{\mathrm{SO}} t^3, \label{eq:U_U_SO}
\end{equation}
where $t$ is the time step and $W_{\mathrm{SO}}$ is the worst-case Trotter error constant of the SO scheme. This Trotter error constant can be evaluated using the following nested-commutator expression \cite{suzuki_1985, Kivlichan2020ImprovedTrotterization, Childs2021TheoryScaling}
\begin{equation}
     W_{\mathrm{SO}} = \frac{1}{24} \Big\lVert [[V, T], V] \Big\rVert + \frac{1}{12} \Big\lVert [[V, T], T] \Big\rVert.
     \label{eq:W_SO}
\end{equation}
We refer to the nested commutators $[[V,T],V]$ and $[[V,T],T]$ as $\OVTV$ and $\OVTT$ respectively. The spectral norm of these nested commutators is most commonly upper-bounded using the triangle inequality. However, depending on how this is performed, such an approach can further loosen the worst-case Trotter error significantly. Instead, Ref.~\cite{blunt2025montecarloapproachbound} introduced an alternative approach based on projector Monte Carlo simulation. This method makes use of the following fact: if we consider a matrix $O$ with elements $O_{ij}$, and a matrix $\abs(O)$ with elements $|O_{ij}|$, then
\begin{equation}
    \lVert O \rVert \le \lVert \abs(O) \rVert.
    \label{eq:monte_carlo_upper_bound}
\end{equation}
If $O$ is a Hermitian matrix, as is the case for both $\OVTV$ and $\OVTT$, then the spectral norm of $\abs(O)$ is equal to the largest eigenvalue of $\abs(O)$, or equivalently to minus the lowest eigenvalue of $-\abs(O)$. This is valuable because $-\abs(O)$ is a sign-problem-free matrix, hence the lowest eigenvalue of $-\abs(O)$ can be obtained efficiently up to large system sizes by projector Monte Carlo simulation. It was shown in Ref.~\cite{blunt2025montecarloapproachbound} that Eq.~\eqref{eq:monte_carlo_upper_bound} is typically a very tight upper bound for the particular second-order Trotter error commutators studied, which are closely-related to those investigated in this paper. In particular, we find that the upper bound on $\lVert \OVTV \rVert$ is essentially exact. We refer to Ref.~\cite{blunt2025montecarloapproachbound} for a detailed description of this method.

In this paper, we sample $\OVTV$ and $\OVTT$ from a slightly different version of the PPP model than that considered in Ref.~\cite{blunt2025montecarloapproachbound}. We show the worst-case bounds of acenes, rhombenes and triangulenes in Fig.~\ref{fig:Commutator_bounds}, and fit the data to a power law to provide approximate system size scalings.

Similarly, we define the worst-case error between the exact time evolution operator, $\mathcal{U}$, and its second-order approximation, $U_{\mathrm{tile}}$, as
\begin{equation}
   \lVert \mathcal{U} - U_{\mathrm{tile}}(t) \rVert \leq W_{\mathrm{tile}} t^3,
\end{equation}
As shown in Ref.~\cite{Bay-Smidt2025}, this Trotter error constant can be upper bounded by
\begin{equation}
    W_{\mathrm{tile}} \leq W_{\mathrm{SO}} + W_{T},
\end{equation}
where $W_{\mathrm{SO}}$ can be evaluated using Eq.~\eqref{eq:W_SO} and $W_T$ is defined from the spectral norm difference between the exact time evolution of the kinetic energy operator, $\mathcal{U}_T =e^{-iTt}$ and its second order approximation, Eq.~\eqref{eq:U_T},
\begin{equation}
    \lVert \mathcal{U}_T - U_T \rVert \leq W_T t^3.
\end{equation}
We explain in Section~\ref{app:free_fermionic_operators_error} how we obtain tight bounds on $W_T$, and we plot results for acenes, rhombenes and triangulenes in Fig.~\ref{fig:Kinetic_Trotter_error}.

\subsection{Average-case error \label{app:average_case_error}}
The average-case Trotter error between the exact Hamiltonian simulation unitary, $\mathcal{U}$, and its second-order split-operator approximation, $U_{\mathrm{SO}}$, is upper bounded by
\begin{equation}
    \mathcal{A}(\mathcal{U},U_{\mathrm{SO}}) \leq \Big[\mathbb{E}_{i \in {\mathcal{E}}}  \lVert ( \mathcal{U} - U_{\mathrm{SO}}) \ket{i} \rVert_2^2 \Big ]^{1/2} \leq A_{\mathrm{SO}} t^3.
\end{equation}
The average-case SO error constant, $A_{\mathrm{SO}}$, can be evaluated through nested-commutator expressions \cite{Zhao2022}
\begin{equation}
    A_{\mathrm{SO}} = \frac{1}{24 \sqrt{d_\mathcal{E}}} \lVert [[V, T], V] \rVert_{F,\mathcal{E}} + \frac{1}{12 \sqrt{d_\mathcal{E}}} \lVert [[V, T], T] \rVert_{F,\mathcal{E}}, \label{eq:A_SO_Appendix}
\end{equation}
 where $d_\mathcal{E}$ is the dimension of our ensemble subspace and $\lVert \, \cdot \, \rVert_{F,\mathcal{E}}$ is the Frobenius norm restricted to that subspace.

For the systems considered here, the dimension of the full computational space is $d=2^{2N}$, where $N$ is the number of carbon atoms and $2N$ is the number of spin orbitals. However, when performing quantum simulations of chemical systems, we are typically interested in states from a computational subspace with a fixed number of electrons. For the carbon-based nanographenes, we initialize $N$ electrons in the $2N$ spin orbitals, reducing the dimension of the relevant computational space to $d_N=\binom{2N}{N}$. We further restrict this subspace to the $S_z=0$ symmetry sector for systems with an even number of electrons (acenes, rhombenes, 3-triangulene and 5-triangulene) and the $S_z=1/2$ subspace for systems with an odd number of electrons (2-triangulene and 4-triangulene). This reduces the subspace dimensions to $d_{S_z = 0} = \binom{N}{N/2}^2$ and $d_{S_z = 1/2} = \binom{N}{(N-1)/2}\binom{N}{(N+1)/2}$.

Our goal is to estimate the normalized Frobenius norm of an operator, $O$, over a relevant computational subspace with certain particle number and spin symmetries of a given operator. This subspace Frobenius norm is defined as
\begin{equation}
 \frac{1}{\sqrt{d_{\mathcal{E}}}}\lVert O \rVert_{F,\mathcal{E}} = \frac{1}{\sqrt{d_{\mathcal{E}}}} \sqrt{ \sum_{\ket{i} \in \mathcal{E}} \bra{i}O^\dagger O \ket{i}},
\end{equation}
and we specifically consider the operators $\OVTV$ or $\OVTT$ in this paper. This expression can be written as
\begin{equation}
     \frac{1}{\sqrt{d_{\mathcal{E}}}}\lVert O \rVert_{F,\mathcal{E}} = \frac{1}{\sqrt{d_{\mathcal{E}}}} \sqrt{ \sum_{\ket{i}  \in \mathcal{E}}  \lVert O\ket{i} \rVert_2^2},
\end{equation}
where $\{ | i \rangle \}$ are computational basis states in the subspace $\mathcal{E}$. We estimate the normalized Frobenius norm by sampling over states within the subspace. Using $K$ states sampled uniformly from the relevant subspace, we can estimate the normalized Frobenius norm as
\begin{equation}
    \frac{1}{\sqrt{d_{\mathcal{E}}}}\lVert O \rVert_{F,\mathcal{E}}  \approx \frac{1}{\sqrt{K}} \sqrt{\sum_{k}^K  \lVert O\ket{i_k} \rVert_2^2} \;, \; \; \; | i_k \rangle \in \mathcal{E}, \label{eq:norm_frob_norm_sampling}
\end{equation}
with large enough $K$. The subscript $k$ is a label on the uniformly sampled states $\ket{i}$. We calculated this norm estimate of $\OVTV$ and $\OVTT$ for the different classes of nanographenes, and the results are shown in Fig.~\ref{fig:Commutator_bounds}. Note that we never construct the $\OVTV$ and $\OVTT$ matrices explicitly as this quickly becomes intractable. Instead, we construct the nested commutators in the Pauli representation. Then, we apply this Pauli operator to the computational basis states $\ket{i}$, which is computationally efficient, and calculate the resulting 2-norm. These nested commutator norm estimates allow us to calculate $A_{\mathrm{SO}}$ (the average-case error constant of $U_{\mathrm{SO}}$) using Eq.~\eqref{eq:A_SO_Appendix}.

Next, we consider the average-case Trotter error of $U_{\mathrm{tile}}$. The average-case Trotter error between $\mathcal{U}$ and $U_{\mathrm{tile}}$ is upper bounded by
\begin{equation}
     \mathcal{A}(\mathcal{U},U_{\mathrm{tile}}) \leq \Big[ \mathbb{E}_{i \in {\mathcal{E}}} \lVert ( \mathcal{U} - U_{\mathrm{tile}}) \ket{i} \rVert_2^2 \Big ]^{1/2} \leq A_{\mathrm{tile}} t^3,
\end{equation}
which can be written as \cite{Zhao2022}
\begin{equation}
     \Big[ \mathbb{E}_{i \in {\mathcal{E}}} \lVert ( \mathcal{U} - U_{\mathrm{tile}}) \ket{i} \rVert_2^2 \Big ]^{1/2} = \frac{1}{\sqrt{d_{\mathcal{E}}}} \lVert \mathcal{U} - U_{\mathrm{tile}} \rVert_{F,\mathcal{E}} \;.
\end{equation}
We show that the tile Trotterization average-case norm is upper bounded by $A_{\mathrm{tile}} \leq A_{\mathrm{SO}} + A_T$, by proving that
\begin{equation}
    \lVert \mathcal{U} -  U_{\mathrm{tile}} \rVert_{F,\mathcal{E}} \leq \lVert \mathcal{U} -  U_{\mathrm{SO}} \rVert_{F,\mathcal{E}} + \lVert \mathcal{U}_T -  U_{T} \rVert_{F,\mathcal{E}}. \label{eq:L2_norm_proof}
\end{equation}
We start by rewriting $\lVert \mathcal{U} -  U_{\mathrm{tile}}\rVert_{F,\mathcal{E}}$ and applying the triangle inequality
\begin{equation}
       \! \! \! \! \! \!  \! \! \!  \! \lVert  \mathcal{U} - U_{\mathrm{SO}} + U_{\mathrm{SO}} - U_{\mathrm{tile}} \rVert_{F,\mathcal{E}} \leq \lVert \mathcal{U} - U_{\mathrm{SO}}\rVert_{F,\mathcal{E}} + \lVert U_{\mathrm{SO}} - U_{\mathrm{tile}} \rVert_{F,\mathcal{E}} . \label{eq:2_norm_proof_1}
\end{equation}
The norm $\lVert U_{\mathrm{SO}} - U_{\mathrm{tile}} \rVert_{F,\mathcal{E}}$ can be rewritten using Eqs.~\eqref{eq:U_SO_app} and \eqref{eq:U_SO_tile_app_v2} and the unitary invariance of the Frobenius norm
\begin{equation}
    \lVert U_{\mathrm{SO}} - U_{\mathrm{tile}} \rVert_{F,\mathcal{E}} = \lVert e^{-iVt/2} (\mathcal{U}_T - U_{T}) e^{-iVt/2} \rVert_{F,\mathcal{E}} = \lVert\mathcal{U}_T - U_{T}\rVert_{F,\mathcal{E}}. \label{eq:2_norm_proof_2}
\end{equation}

Using Eqs.~\eqref{eq:2_norm_proof_1}--\eqref{eq:2_norm_proof_2}, we can show that Eq.~\eqref{eq:L2_norm_proof} is true. Then the average-case error estimates of $U_{\mathrm{tile}}$ can be upper bounded by
\begin{equation}
 \frac{1}{\sqrt{d_{\mathcal{E}}}} \lVert \mathcal{U} - U_{\mathrm{tile}} \rVert_{F,\mathcal{E}} \leq A_{\mathrm{tile}} t^3 \leq (A_{\mathrm{SO}} + A_T )t^3,
\end{equation}
where $A_{\mathrm{SO}}$ can be computed using the nested commutator expression in Eq.~\eqref{eq:A_SO_Appendix} and $A_T$ can be evaluated from
\begin{equation}
 \frac{1}{\sqrt{d_{\mathcal{E}}}} \lVert\mathcal{U}_T- U_T \rVert_{F, \mathcal{E}} \leq A_{T} t^3.
\end{equation}
We show how to evaluate $A_T$ in Appendix~\ref{app:free_fermionic_operators_error}, and present numerical results for acenes, rhombenes and triangulenes in Fig.~\ref{fig:Kinetic_Trotter_error}.

\subsection{Free fermionic operators \label{app:free_fermionic_operators_error}}
The Trotter error of free fermionic operators can be bounded efficiently through nested-commutator expressions as shown in Ref.~\cite{Campbell_2022}. However, here we show that such free fermionic worst-case error bounds can be further tightened by exploiting dimensionality reduction and furthermore show how to use our approach to sample average-case error constants of free fermionic operators.

We consider a free fermionic operator, $T$, as a sum of $S$ free fermionic operators, so $T=\sum_{s=1}^S T_s$. Each $T_s$ acts on $n$ modes, defined as
\begin{equation}
    T_{s} = \sum_{i,j=1}^n [A_s]_{i,j} a_i^\dagger a_j,
\end{equation}
leading to
\begin{equation}
    T = \sum_{s=1}^S T_s = \sum_{i,j=1}^n [A]_{i,j} a_i^\dagger a_j,
\end{equation}
where $A = \sum_{s}^S A_s$, and all matrices $A$ have zeroes on the diagonal ($[A]_{j,j} = 0$ for all $j$). We define the exact unitary time evolution operator of $T$ as
\begin{equation}
    \mathcal{U}_T = e^{-iTt} = e^{-i (\sum_{s=1}^S T_s) t},
\end{equation}
and its second-order approximation
\begin{equation}
    U_T = \prod_{s=1}^S e^{-iT_s t/2} \prod_{s=S}^1 e^{-iT_s t/2}.
\end{equation}

Before considering the worst- and average-case error bounds for Trotterized free fermionic Hamiltonians, we present two Lemmas on dimensionality reduction.
\begin{lemma}[Norm Correspondence] \label{lem:norm_correspondence}
For a free fermionic Hamiltonian $H = \sum_{ij} [A]_{ij} a_i^\dagger a_j$ with zeroes on the diagonal, the operator (a.k.a. spectral) norm on the Fock space is related to the trace norm of the underlying matrix $A$ by
\begin{equation}
    \lVert H \rVert = \frac{1}{2} \lVert A \rVert_{tr},
\end{equation}
where $\lVert A \rVert_{tr} =\sum_i \sigma_i(A)$ is the sum of the singular values of $A$.
\end{lemma}
\noindent For a proof see Appendix~A of Ref.~\cite{Campbell_2022}.
\begin{lemma}[Lie Algebra Isomorphism] \label{lem:lie_algebra}
The mapping between the Lie algebra of quadratic fermionic operators and the Lie algebra of $n \times n$ matrices preserves the exponential structure. Specifically, if:
\begin{equation}
     e^{-iA_1 t} e^{-iA_2t} = e^{-i \tilde{A}t}
\end{equation}
then it follows that
\begin{equation}
    e^{-iT_1t} e^{-iT_2 t} = e^{-i \tilde{T}t}
\end{equation}
where $T_k = \sum_{ij} [A_k]_{ij} a_i^\dagger a_j$ and $\tilde{T} = \sum_{ij} [\tilde{A}]_{ij} a_i^\dagger a_j$.
\end{lemma}
\noindent While this second lemma is also well-known folklore, we offer a self contained proof:

\begin{proof}
The proof proceeds in two steps. First, we establish that the mapping from matrices to quadratic operators is a Lie algebra homomorphism (i.e., it strictly preserves the commutator). Second, we use the Baker-Campbell-Hausdorff formula to lift this homomorphism to the corresponding Lie groups.

Step 1: The Lie Algebra Homomorphism. Let $A$ be an $n \times n$ matrix. We define a linear map $T$ to the fermionic Fock space as:
\begin{equation}
    T(A) = \sum_{i,j=1}^{n} A_{ij} a_i^\dagger a_j
\end{equation}
where $a_i^\dagger$ and $a_j$ obey the canonical anticommutation relations (CAR):
\begin{equation}
    \{a_i, a_j^\dagger\} = \delta_{ij}, \quad \{a_i^\dagger, a_j^\dagger\} = \{a_i, a_j\} = 0.
\end{equation}
We wish to evaluate the commutator $[T(A), T(B)]$ for two arbitrary matrices $A$ and $B$:
\begin{equation}
    [T(A), T(B)] = \sum_{i,j,k,l} A_{ij} B_{kl} [a_i^\dagger a_j, a_k^\dagger a_l].
\end{equation}
Using the standard operator identity
\begin{align}
[WX, YZ] = & W\{X, Y\}Z - WY\{X, Z\} + \{W, Y\}ZX - Y\{W, Z\}X
\end{align}
and setting $W=a_i^\dagger$, $X=a_j$, $Y= a_k^\dagger$ and $Z=a_l$, we find that $\{X,Z\}=0$ and $\{W,Y \}=0$, so that:
\begin{align}
    [a_i^\dagger a_j, a_k^\dagger a_l] &= a_i^\dagger \{a_j, a_k^\dagger\} a_l - a_k^\dagger \{a_i^\dagger, a_l\} a_j, \\ \nonumber
    &= a_i^\dagger (\delta_{jk}) a_l - a_k^\dagger (\delta_{il}) a_j.
\end{align}
Substituting this back into the sum yields:
\begin{equation}
    [T(A), T(B)] = \sum_{i,j,k,l} A_{ij} B_{kl} \left( \delta_{jk} a_i^\dagger a_l - \delta_{il} a_k^\dagger a_j \right).
\end{equation}
Evaluating the Kronecker deltas collapses the sums over $k$ and $l$ in the respective terms:
\begin{equation}
    [T(A), T(B)] = \sum_{i,j,l} A_{ij} B_{jl} a_i^\dagger a_l - \sum_{i,j,k} B_{ki} A_{ij} a_k^\dagger a_j.
\end{equation}
By renaming the dummy indices in the second term ($k \to i$, $i \to j$, $j \to l$), we obtain:
\begin{align}
    [T(A), T(B)] &= \sum_{i,l} (AB)_{il} a_i^\dagger a_l - \sum_{i,l} (BA)_{il} a_i^\dagger a_l \\
    &= \sum_{i,l} (AB - BA)_{il} a_i^\dagger a_l \\
    &= \sum_{i,l} [A, B]_{il} a_i^\dagger a_l.
\end{align}
Thus, we have established the exact homomorphism:
\begin{equation}
    [T(A), T(B)] = T([A, B]).
\end{equation}

Step 2: Lifting to the Lie Group. The Baker--Campbell--Hausdorff (BCH) theorem states that for elements $X$ and $Y$ in a Lie algebra, the logarithm of the product of their exponentials can be expressed entirely in terms of nested commutators:
\begin{align}
    \ln(e^X e^Y) =& X + Y + \frac{1}{2}[X, Y] + \frac{1}{12}[X, [X, Y]] - \frac{1}{12}[Y, [X, Y]] + \dots
\end{align}
Given the matrix relation:
\begin{equation}
    e^{-i A_1 t} e^{-i A_2 t} = e^{-i \tilde{A} t}
\end{equation}
the exponent $-i \tilde{A} t$ is exactly the BCH series evaluated for $X = -i A_1 t$ and $Y = -i A_2 t$.

Now consider the operator product $e^{-i T_1 t} e^{-i T_2 t}$, where $T_k = T(A_k)$. Because our mapping strictly preserves commutators (i.e., $[T(X), T(Y)] = T([X, Y])$), every nested commutator in the BCH series for the operators evaluates exactly to the mapping of the corresponding matrix commutator.

By linearity, the entire infinite BCH series for the operators collapses to $T$ applied to the matrix BCH series:
\begin{equation}
    \ln \left( e^{T(-i A_1 t )} e^{T(-i A_2 t )} \right) = T(-i \tilde{A} t).
\end{equation}
Exponentiating both sides yields the final result:
\begin{equation}
    e^{-i T_1 t} e^{-i T_2 t} = e^{-i \tilde{T} t},
\end{equation}
where $\tilde{T} = T(\tilde{A})$, completing the proof.
\end{proof}

\begin{figure*}[t]
    \centering
    \includegraphics[width=1.0\linewidth]{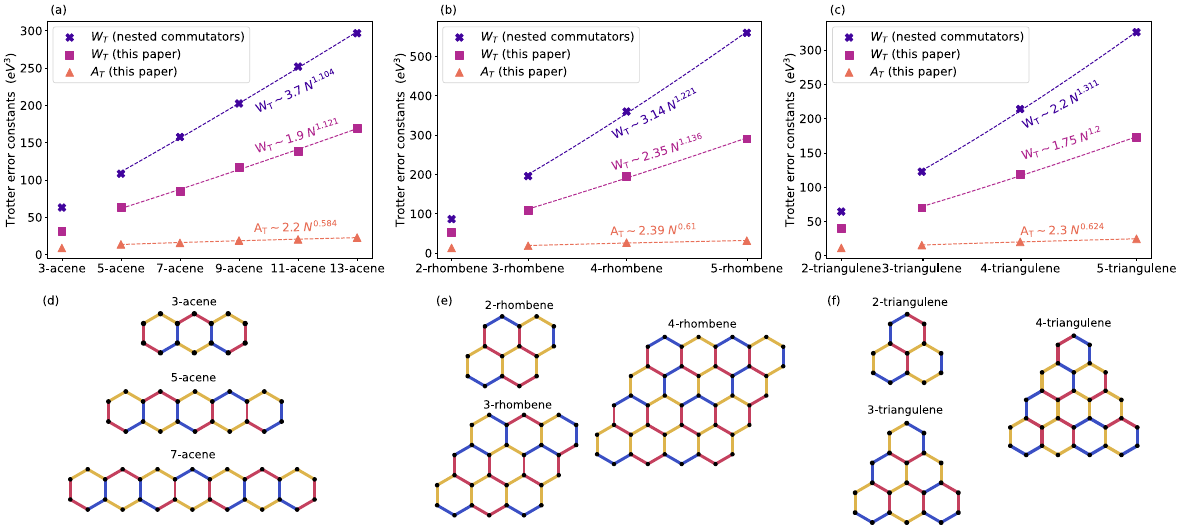}
    \caption{Worst- and average-case Trotter errors from the decomposition of the kinetic energy operator of (a) acenes, (b) rhombenes and (c) triangulenes. The plots show a comparison between the worst-case error constant, $W_T$, evaluated using nested commutators as in Refs.~\cite{Campbell_2022} and \cite{Bay-Smidt2025} to $W_T$, evaluated using the free fermionic operator norm strategy outlined in this paper (Section~\ref{sec:W_T}). The approach presented in this paper gives tighter bounds on the Trotter error constant and avoids construction of potentially many nested commutators (see Appendix~C of Ref.~\cite{Bay-Smidt2025}). The average-case error constant is evaluated using the strategy presented in Appendix~\ref{app:free_fermion_avg_case}. We show the splitting of the kinetic energy operator terms into different sections (red, blue and gold) in (d), (e) and (f), that are time-evolved separately -- meaning that for these systems, the time evolution of the kinetic energy operator is constructed as $U_T = e^{-iT_{\mathrm{red}} \frac{t}{2}} e^{-iT_{\mathrm{blue}}\frac{t}{2}} e^{-iT_{\mathrm{gold}}t}  e^{-iT_{\mathrm{blue}}\frac{t}{2}}  e^{-iT_{\mathrm{red}}\frac{t}{2}} $. We refer to Ref.~\cite{Bay-Smidt2025} for further details on how each kinetic energy section, $T_s$, can be recovered from those plots. The underlying matrices, $A$, of each section (red, blue, gold) that we have used in this paper, can be generated from code made available on Zenodo (see data availability section). These calculations have been performed at half-filling ($N$ electrons in $2N$ spin orbitals) and in the $S_z = 0$ subspace for all systems except for 2- and 4-triangulene where calculations were performed in the $S_z = 1/2$ subspace.}
    \label{fig:Kinetic_Trotter_error}
\end{figure*}

We use the definitions and Lemmas presented above to establish methods for evaluating worst- and average-case error bounds for free fermionic operator splittings. In Fig.~\ref{fig:Kinetic_Trotter_error}, we present the numerical results for the worst- and average-case Trotter error constants of the acenes, rhombenes and triangulenes considered in this paper. We also show the corresponding error bounds obtained through nested-commutator bounds as in Refs.~\cite{Campbell_2022, Childs2021TheoryScaling, Bay-Smidt2025}, that are looser by approximately a factor two. The kinetic energy operator sections are constructed using the tile-Trotterization scheme \cite{Bay-Smidt2025}, and we show kinetic operator ``sections'' ($T_s$) of a subset of the molecules considered in this paper in Fig.~\ref{fig:Kinetic_Trotter_error}. We also include code (see data availability statement or Ref.~\cite{zenodo_data}) where the underlying adjacency matrices of each kinetic energy operator section can be generated and used directly to calculate our free fermionic operator norm results.

\subsubsection{Worst-case free fermionic operator error\label{sec:W_T}}
Our goal is to bound the spectral norm
\begin{equation}
    \lVert \mathcal{U}_T-U_T \rVert \leq W_T t^3, \label{eq:free_fermion_spectral_norm}
\end{equation}
and determine the Trotter error constant $W_T$. By the unitary invariance of the spectral norm, we have
\begin{equation}
    \lVert \mathcal{U}_T-U_T \rVert = \lVert \one - \mathcal{U}_T^{\dagger} U_T \rVert, \label{eq:free_ferm_spectral_norm}
\end{equation}
The product $\mathcal{U}_T^{\dagger} U_T$ is given by
\begin{equation}
     \mathcal{U}_T^{\dagger} U_T = e^{iTt}\prod_{s=1}^S e^{-iT_s t/2} \prod_{s=S}^1 e^{-iT_s t/2} =e^{-i \tilde{T}_{\Delta} t}, \label{eq:UdagU_wc_derivation}
\end{equation}
where $\tilde{T}_{\Delta}$ is the effective Hamiltonian responsible for the unitary evolution $\mathcal{U}_T^{\dagger} U_T$. We can write Eq.~\eqref{eq:free_ferm_spectral_norm} as
\begin{align}
     \lVert \one - \mathcal{U}_T^{\dagger} U_T \rVert & = \lVert \one - e^{-i\tilde{T}_\Delta t} \rVert
     = \mathrm{max}_{\lambda \in \Lambda(\tilde{T}_\Delta )} \lvert 1- e^{-i \lambda t} \rvert, \label{eq:UdagU_norm_max_eigenvalue}
\end{align}
where $\Lambda(\tilde{T}_\Delta)$ denotes the set of eigenvalues of $\tilde{T}_\Delta$. Since the spectrum is symmetric about zero (a consequence of the underlying $A$ matrix having zeroes on the diagonal), we need only the positive eigenvalues. Furthermore, since $| 1- e^{-i \lambda t} |$ is monotonically increasing on the interval $\lambda t \in [0, \pi]$ it follows that provided $||\tilde{T}_\Delta ||\, t \leq \pi$ the maximum is achieved by the largest eigenvalue and so
\begin{align}
     \lVert \one - \mathcal{U}_T^{\dagger} U_T \rVert & = \lvert 1- e^{-i || \tilde{T}_\Delta || t} \rvert.\label{eq:UdagU_norm_evaluation}
\end{align}
Using Lemma~\ref{lem:lie_algebra}, Eq.~\eqref{eq:UdagU_wc_derivation} can be written in terms of the underlying $n \times n$ matrices
\begin{equation}
     e^{iAt} \prod_{s=1}^S e^{-iA_s t/2} \prod_{s=S}^1 e^{-iA_s t/2} = e^{-i \tilde{A}_\Delta t},
\end{equation}
where $\tilde{A}_\Delta$ is the underlying matrix of $\tilde{T}_\Delta$.
We find $\tilde{A}_\Delta$ as
\begin{equation}
     \tilde{A}_\Delta = \frac{i}{t} \log(e^{-i \tilde{A}_\Delta t}),
\end{equation}
and use Lemma~\ref{lem:norm_correspondence} to evaluate the spectral norm of $\tilde{T}_\Delta$
\begin{equation}
     \lVert \tilde{T}_\Delta \rVert = \frac{1}{2} \lVert \tilde{A}_\Delta \rVert_{tr}. \label{eq:spectral_norm_trace_norm}
\end{equation}
In practice, we evaluate this norm for a range of relevant time steps, $0.01 \; \mathrm{eV}^{-1} \leq t \leq 0.05 \; \mathrm{eV}^{-1}$, insert them into Eq.~\eqref{eq:UdagU_norm_evaluation}, and fit the Trotter error constant, $W_T$, from Eq.~\eqref{eq:free_fermion_spectral_norm}. For the type of free fermionic operators considered here, this procedure calculates the spectral norm in the half-filling subspace which is also the worst-case spectral norm across all particle-number sectors (see Appendix~A of Ref.~\cite{Campbell_2022}).

In some cases, for example when considering free fermionic operators away from half-filling and for the average-case errors, it is convenient to write $\tilde{A}_\Delta$ as
\begin{equation}
     \tilde{A}_\Delta = \sum_{j} \lambda_j \ket{j}\bra{j},
\end{equation}
where $\lambda_j$ is the energy of mode $j$. Within a given subspace $\mathcal{E}$, for example the $N$-particle subspace, the spectral norm, $\lVert \tilde{T}_\Delta \rVert$, can be found by summing the $N$ eigenmodes $\lambda_j$ that give the largest absolute value. For the free fermionic operators considered here, the eigenmodes $\lambda_j$ are distributed evenly around zero \cite{Campbell_2022}. Let $\vec{\lambda}$ be a vector of eigenmodes in descending order such that $\lambda_1 \geq \lambda_2, ...  \geq\lambda_n$, then
\begin{equation}
     \lVert \tilde{T}_\Delta \rVert = \sum_{j=1}^N \lambda_j \; .\label{eq:spectral_norm_eigenspectrum}
\end{equation}
At half-filling, $N = n/2$, this is equivalent to Eq.~\eqref{eq:spectral_norm_trace_norm}, but away from half filling, Eq.~\eqref{eq:spectral_norm_eigenspectrum} gives lower (tighter) bounds (see e.g. \cite{su2021nearly,mcardle2022exploiting}). Spin-symmetry can also be included by requiring a certain number of particles within the spin-up and spin-down sectors. For hopping Hamiltonians, there is no coupling between the spin-up and spin-down sectors, allowing us to treat each sector independently. The derivations above can therefore also be viewed as being applied to a single spin-sector at a time (spin-up or spin-down) and the total norm can be obtained by summing the norm contribution from each spin-sector.

\subsubsection{Average-case free fermionic operator error \label{app:free_fermion_avg_case}}

The average-case error contribution can be obtained from the Frobenius norm
\begin{equation}
        \frac{1}{\sqrt{d_{\mathcal{E}}}}|| \mathcal{U}_T - U_{T} ||_{F,\mathcal{E}} \leq A_T t^3.
        \label{eq:avg_case_free_fermion_error}
\end{equation}
Our goal is to evaluate this expression and determine the average-case error constant, $A_T$. We use the unitary invariance of the Frobenius norm
\begin{equation}
        \frac{1}{\sqrt{d_{\mathcal{E}}}}|| \mathcal{U}_T - U_{T} ||_{F,\mathcal{E}} = \frac{1}{\sqrt{d_{\mathcal{E}}}}  || \one - e^{-i\tilde{T}_\Delta t} ||_{F,\mathcal{E}} \; .
\end{equation}
Using the standard Frobenius norm equality $|| X ||_{F,\mathcal{E}} = \sqrt{ \mathrm{Tr_{\mathcal{E}}}(X^\dagger X)}$ for any $X$, we have
\begin{align}
    || \mathcal{U}_T - U_{T} ||_{F,\mathcal{E}}& = \sqrt{ \mathrm{Tr}_\mathcal{E}[(\one - e^{i\tilde{T}_\Delta t})(\one - e^{-i\tilde{T}_\Delta t})  ]  } = \sqrt{ 2d_\mathcal{E} - 2 \mathrm{Re}( \mathrm{Tr_\mathcal{E}}[ e^{i\tilde{T}_\Delta t} ] )} \; ,
\end{align}
leading to
\begin{equation}
        \frac{1}{\sqrt{d_{\mathcal{E}}}}|| \mathcal{U}_T - U_{T} ||_{F,\mathcal{E}} = \sqrt{ 2 - 2 \mathrm{Re}( \mathrm{Tr_\mathcal{E}}[ e^{i\tilde{T}_\Delta t} ] / d_\mathcal{E})}.
        \label{eq:avg_case_free_fermion_error_computation}
\end{equation}
Evaluating $\mathrm{Tr_\mathcal{E}}[ e^{i\tilde{T}_\Delta t}] /d_\mathcal{E}$ exactly requires taking the trace over an exponentially large subspace. Similarly to the average-case SO errors, we employ a sample-based algorithm to get around this. Specifically, we let $\vec{\lambda}$ be a vector denoting the energies associated with the eigenmodes of $\tilde{T}_\Delta$, so that $\tilde{T}_\Delta \ket{\vec{x}} = (\vec{\lambda}\cdot \vec{x})\ket{\vec{x}}$ and consequently $e^{i\tilde{T}_\Delta t}  \ket{ \vec{x}} = e^{i \vec{\lambda}\cdot \vec{x} t}  \ket{ \vec{x}}$. Recall that the elements of $\vec{\lambda}$ can be efficiently classically computed as they are the eigenvalues of the $n$-by-$n$ matrix $\tilde{A}_\Delta$. Using $K$ states sampled uniformly from the relevant subspace, the normalized trace in Eq.~\eqref{eq:avg_case_free_fermion_error_computation} can be estimated as
\begin{equation}
    \mathrm{Tr_\mathcal{E}}[ e^{i\tilde{T}_\Delta t} ] / d_\mathcal{E} \approx \frac{1}{K} \sum_{k}^K e^{i \vec{\lambda} \cdot \vec{x_k} t} , \; \; \vec{x_k} \in \mathcal{E},
    \label{eq:avg_case_free_fermion_sampling}
\end{equation}
where $k$ is a label on the uniformly sampled states $\vec{x}$.
Combining Eqs.~\eqref{eq:avg_case_free_fermion_error_computation} and \eqref{eq:avg_case_free_fermion_sampling}, we can estimate the average-case error contribution from the type of kinetic energy operator splittings considered here for a range of time steps ($0.01 \; \mathrm{eV}^{-1} \leq t \leq 0.05 \; \mathrm{eV}^{-1} $), and evaluate the average-case error constant $A_T$ by fitting to Eq.~\eqref{eq:avg_case_free_fermion_error}.

\subsection{Energy and gap errors \label{App:phase_error}}
The PPP Hamiltonian is defined as $H=T+V$, with $T$ and $V$ given by Eq.~\eqref{eq:T} and Eq.~\eqref{eq:V}, respectively. The effective Hamiltonian is defined based on the Hamiltonian simulation approximation, $U$, as
\begin{equation}
    U = e^{-i\tilde{H}t},
\end{equation}
and we consider the two Trotter schemes $U_{\mathrm{SO}}$ (Eq.~\eqref{eq:U_SO}) and $U_{\mathrm{tile}}$ (Eq.~\eqref{eq:U_SO_tile}) as our Hamiltonian simulation approximations. When performing Trotterized Hamiltonian simulation, the estimated energies will correspond to the eigenvalues of the effective Hamiltonian $\tilde{H}$ instead of the original Hamiltonian $H$. Therefore, for QPE, we need to estimate the error in the eigenvalues of the Trotterized Hamiltonian.

For second-order Trotter schemes, the error on the energy of eigenstate $m$, denoted $E_m$, can be used to define an error constant, $C_m$, by
\begin{equation}
    \lvert E_{m} - \Tilde{E}_m \rvert = C_m t^2. \label{eq:absolute_phase_error_app}
\end{equation}
Note that $C_m$ depends on the Trotter time step size, $t$. We do not indicate this explicitly but it is an important distinction between energy errors and  worst- and average-case error constants, which are independent of $t$. These quantities can be evaluated exactly for small systems. For such small systems we can diagonalize $H$ to obtain $E_m$. To find $\tilde{E}_m$, one can obtain $\tilde{H}$ by
\begin{equation}
    \tilde{H} = \frac{i}{t} \log(U),
    \label{eq:h_eff_def}
\end{equation}
and calculate $\tilde{E}_m$ by diagonalizing $\tilde{H}$.

We also consider calculating energy gaps between two states, $m$ and $n$. We define the exact energy gap between $m$ and $n$ as
\begin{equation}
    \delta_{m,n} = E_m - E_n
\end{equation}
and the corresponding effective energy gap as
\begin{equation}
    \tilde{\delta}_{m,n} = \tilde{E}_m - \tilde{E}_n.
\end{equation}
This allows us to define the gap error constant, $C_{m,n}$ from
\begin{equation}
    \lvert \delta_{m,n} - \Tilde{\delta}_{m,n} \rvert = C_{m,n} t^2,
\end{equation}
which again is defined for a fixed time step, $t$. In our analysis, we use this error differently than the other errors since we do not expect the gap errors, $ \lvert \delta_{m,n} - \Tilde{\delta}_{m,n} \rvert$ to simply scale by some constant times $t^2$ for a range of time step sizes, as we do for the worst-case, average-case and energy errors. However, defining the Trotter error constant is still useful for comparing different error constants at fixed $t$.

We next describe a novel numerical approach to estimate Trotter error in eigenvalues of $\tilde{H}$ for much larger systems than can be considered by exact diagonalization.

\subsubsection{A tensor network approach to calculate Trotter error in eigenvalues}

For a given Trotter step size, $t$, we wish to calculate $E_m - \tilde{E}_m$ for a given energy state $m$, which in our case will be low-lying eigenstates of a nanographene. We will achieve this task using time-dependent DMRG methods (TD-DMRG). Since we rely on the DMRG algorithm, this numerical approach will only be efficient for 1D, quasi-1D or small 2D systems. This allows us to perform accurate calculations for all of the acenes in this paper, but also the small triangulenes and rhombenes. We find that DMRG performs poorly for the larger 2D nanographenes. Still, this subset of systems allows us to go well beyond the systems that can be studied using brute force construction and diagonalization of $\tilde{H}$.

First, note that we can use DMRG directly to find the desired energies of the true Hamiltonian, $E_m$ (again, assuming a quasi-1D or small 2D nanographene). However, it is more challenging to find the eigenvalues of the effective Hamiltonian, defined by Eq.~\eqref{eq:h_eff_def}, which cannot be easily constructed.

Instead, we take a different approach to calculate $\tilde{E}_m$. We use TD-DMRG to construct a time signal corresponding to the Trotter step, and then perform time series analysis to obtain the desired eigenvalue. Since time evolution is performed with the Trotterized time evolution operator, the eigenvalue obtained will be exactly the eigenvalue of the effective Hamiltonian. The time series analysis follows recent statistical phase estimation approaches, particularly that described in Ref.~\cite{Blunt2023}, which built on similar approaches in Refs.~\cite{Lin2022, Wan2022} and is also closely related to the approach in Ref.~\cite{Wang2023}. Indeed, to obtain $\tilde{E}_m$, we will essentially be performing the statistical phase estimation analysis on data from TD-DMRG. However, since the DMRG results do not have statistical noise as expectation values can be evaluated exactly, it is not correct to call this approach statistical phase estimation, and instead we simply refer to the method as ``time series analysis''.

Given a Trotterized time evolution unitary, $U$, we use TD-DMRG to construct a vector
\begin{equation}
    g_k = \langle \psi | U^k | \psi \rangle.
\end{equation}
Here, $|\psi\rangle$ is an initial state for the Trotter evolution, which should have large overlap with the desired eigenstate whose energy $\tilde{E}_m$ we wish to calculate. We take $| \psi \rangle$ to be the corresponding eigenstate of true Hamiltonian, $H$; this eigenstate can be constructed as a matrix product state with DMRG. Since $\tilde{H}$ is only a small perturbation from $H$, the corresponding eigenstates of the two Hamiltonians will have very high overlap in practice, i.e. $|\langle \Psi_m | \tilde{\Psi}_m \rangle|^2 \approx 1$, where $| \Psi_m\rangle$ and $| \tilde{\Psi}_m\rangle$ are the normalized $m$'th eigenstates of $H$ and $\tilde{H}$, respectively. By using $\one = \sum_m | \tilde{\Psi}_m  \rangle \langle \tilde{\Psi}_m |$, we have
\begin{equation}
    g_k = \sum_n |\langle \psi | \tilde{\Psi}_n \rangle|^2 e^{-i t k \tilde{E}_n}.
    \label{eq:g_k}
\end{equation}
Since $\langle \psi | \tilde{\Psi}_m \rangle \approx 1$, this tells us $g_k \approx e^{-i t k \tilde{E}_m}$. Therefore, it should be simple to estimate $\tilde{E}_m$ to high precision by time series analysis after performing just a few Trotter steps. Still, the sum in Eq.~\eqref{eq:g_k} includes small contributions from other eigenstates, and so it is important to perform a careful analysis which accounts for this. We next describe this general approach to obtain $\tilde{E}_m$ from the vector $g_k$.

We first assume that $\lVert \tilde{H} \rVert t \le \pi/2$, where $\lVert \, \cdot \, \rVert$ denotes the spectral norm. We will explain shortly why this condition can be dropped in practice, but assume it for the following. Following Ref.~\cite{Blunt2023}, we then define a probability distribution associated with $\tilde{H}$,
\begin{equation}
    p(x) = \sum_n | \langle \psi | \tilde{\Psi}_n \rangle|^2 \delta(x - t \tilde{E}_n).
\end{equation}
We then introduce a filter function, $F(x)$, which is some function which has a maximum at $x=0$ and decays rapidly away from this maximum. Then we define
\begin{equation}
    C(x) = \int_{-\pi/2}^{\pi/2} p(y) F(x-y)dy.
    \label{eq:convolution_def}
\end{equation}
Because $C(x)$ is a convolution between $p(x)$ and $F(x)$, it consists of peaks at the locations of eigenvalues of $t\tilde{H}$. Each peak has an amplitude $| \langle \psi | \tilde{\Psi}_n \rangle|^2$. Since $|\langle \psi | \tilde{\Psi}_m \rangle|^2 \approx 1$, we will have $C(x) \approx F(x - t \tilde{E}_m)$, with small contributions from other eigenvalues. Therefore, we can use the location of the maximum of $C(x)$ as an estimate of $t \tilde{E}_m$. By decreasing the width of $F(x)$, this estimate converges to the exact result.

In order to construct $C(x)$, we define $F(x)$ as a truncated Fourier series,
\begin{equation}
    F(x) = \sum_{|k| \le K} F_k e^{ikx}.
    \label{eq:general_fourier_coeffs}
\end{equation}
As a result, $C(x)$ will be a periodic function. We will take $F(x)$ to be an even function, such that $F_{-k} = F_k$. With this Fourier series definition of $F(x)$, a short derivation shows
\begin{align}
    C(x) &= \int_{-\pi/2}^{\pi/2} p(y) F(x-y)dy,\\
    &= \sum_{|k| \le K} F_k e^{ikx} \langle \psi | e^{-it \tilde{H} k} | \psi \rangle,\\
    &= \sum_{|k| \le K} F_k e^{ikx} g_k.
\end{align}
Using $F_{-k} = F_k$, $g_{-k} = g_k^*$ and $g_0 = 1$ we can write
\begin{align}
    C(x) = F_0 + 2\sum_{k=1}^K F_k \Big( \cos(kx) \mathrm{Re}[g_k] - \sin(kx) \mathrm{Im}[g_k] \Big).
    \label{eq:objective_fn_final}
\end{align}
Eq.~\eqref{eq:objective_fn_final} is the final expression that we use to construct $C(x)$, after obtaining $g_k$ from a TD-DMRG calculation using the desired Trotter evolution, and performing $K$ Trotter steps in total.

In our analysis we take $F(x)$ to be a (periodic) Gaussian, and so we take the Fourier coefficients of a Gaussian with width (standard deviation) denoted as $a$. In order to properly resolve the Gaussian, we need $K = \mathcal{O}(1/a)$. Therefore, achieving better resolution requires performing more Trotter steps, as expected. In practice, since we have $|\langle \psi | \tilde{\Psi}_m \rangle|^2 \approx 1$, results are typically converged already using a large value of $a$, or equivalently using only a small value of $K$.

Lastly, note that the above analysis required $\lVert \tilde{H} \rVert t \le \pi/2$. This is formally required to avoid wrapping of the phase, as discussed in Section~\ref{sec:gap_errors} in the context of QPE. However, as also described in the discussion of QPE, this can be relaxed if the spectral weight is sufficiently concentrated around the desired energy. For the setup that we are considering in this section, we again note that $|\langle \psi | \tilde{\Psi}_m \rangle|^2 \approx 1$, and so this criterion is trivially fulfilled. In addition, we can obtain a good estimate of $\tilde{E}_m$ by using $E_m$, calculated by performing DMRG on the original Hamiltonian. Because of these two properties, it is trivial to account for the wrapping of the phase to obtain the desired estimate, $\tilde{E}_m$, should this wrapping occur.

While the above time series analysis gives access to exact $\tilde{E}_m$ for exact $g_k$ and for a sufficient maximum evolution time, in practice $g_k$ as calculated by TD-DMRG will have some error. In particular, the accuracy will depend on the bond dimension (and other parameters) of the performed TD-DMRG simulations. Therefore, results must be converged with respect to the bond dimension, which we find to be possible for quasi-1D or small 2D systems. In the next section we present and discuss our results using this numerical approach.

\begin{figure*}[!t]
    \centering
    \includegraphics[width=0.95\linewidth]{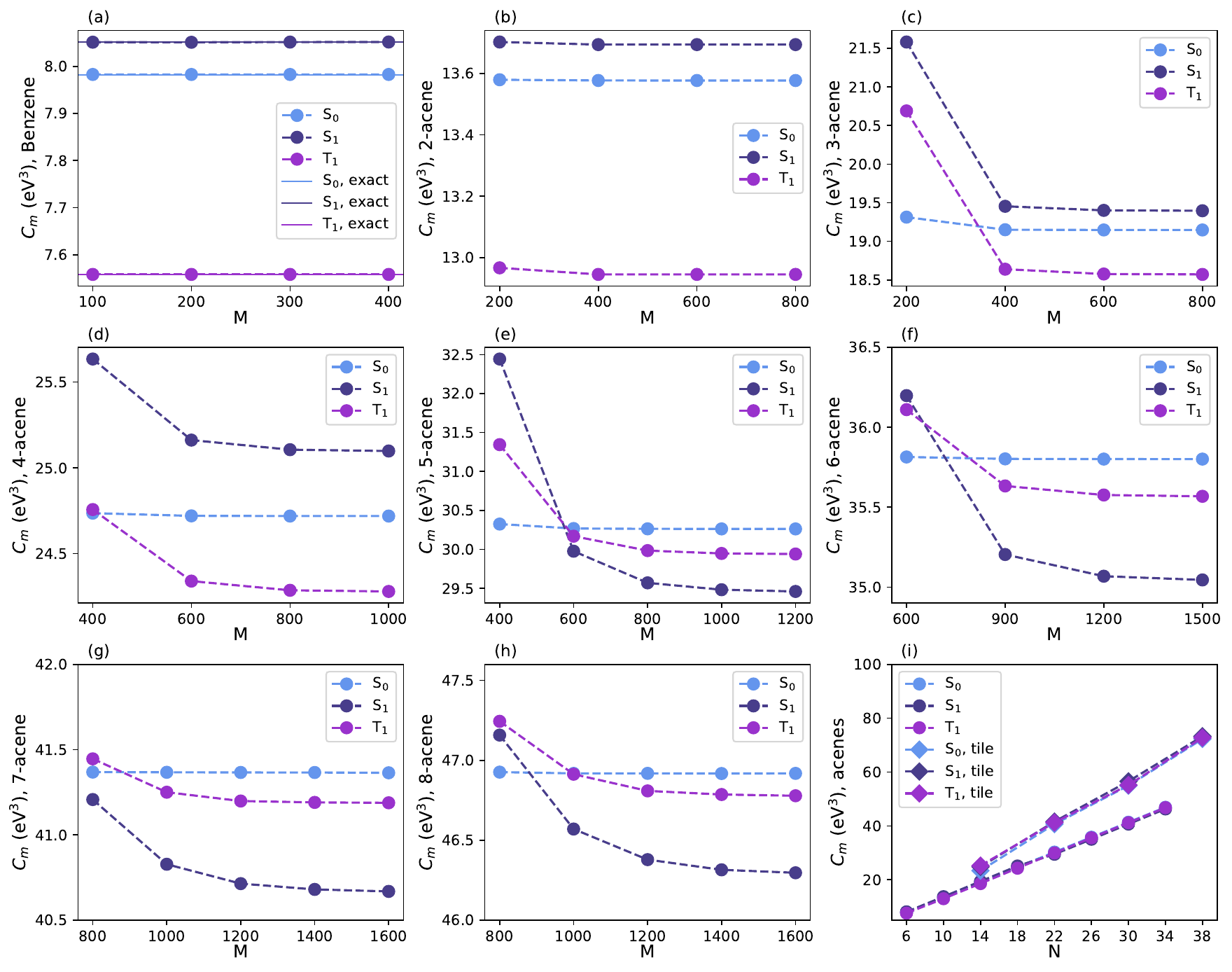}
    \caption{Energy error constant calculations of (a) benzene, (b) 2-acene, (c) 3-acene, (d) 4-acene, (e) 5-acene, (f) 6-acene, (g) 7-acene and (h) 8-acene eigenstates ($\mathrm{S_0}$, $\mathrm{S_1}$ and $\mathrm{T_1}$) of $U_{\mathrm{SO}}$ performed using our TD-DMRG time series analysis with bond dimension $M$. We calculate these energy error constants, $C_m = \lvert E_m-\tilde{E}_m \rvert/t^2$ at $t = 0.01 \; \mathrm{eV}^{-1}$. We show $C_m$ of each system for a range of $M$ and track the convergence of $C_m$ with respect to $M$. For the quasi-1D acenes presented here, the energy error constants converge at relatively low bond dimensions as expected. (i) Shows a comparison of $C_m$ using either $U_{\mathrm{SO}}$ or $U_{\mathrm{tile}}$ as our Hamiltonian simulation unitary as a function of the number of carbon atoms in the acenes. The $U_{\mathrm{tile}}$ values are also computed for $t = 0.01 \; \mathrm{eV}^{-1}$ and using $M = 600$ for 3-acene, $M = 1200$ for 5-acene, $M = 1400$ for 7-acene and $M=1500$ for 9-acene. We clearly see that for same $N$ sizes and for the two respective unitaries, $C_{\mathrm{S_0}} \sim C_{\mathrm{S_1}} \sim C_{\mathrm{T_1}}$, meaning the energy errors on the three low-lying eigenstates of interest are almost equivalent. We fit a power-law to the energy error constants on $\mathrm{S_0}$ of the two respective unitaries, yielding $C_{\mathrm{SO}}=1.34 N^{1.008} \; \mathrm{eV}^3$ (for $n>4$, as in Fig.~\ref{fig:Trotter_errors_acene_rhombene_triangulene}), and $C_{\mathrm{tile}}=1.49 N^{1.066} \; \mathrm{eV}^3$ (for $n>4$).}
    \label{fig:C_estimates_acenes}
\end{figure*}

\begin{figure*}[!t]
    \centering
    \includegraphics[width=0.64\linewidth]{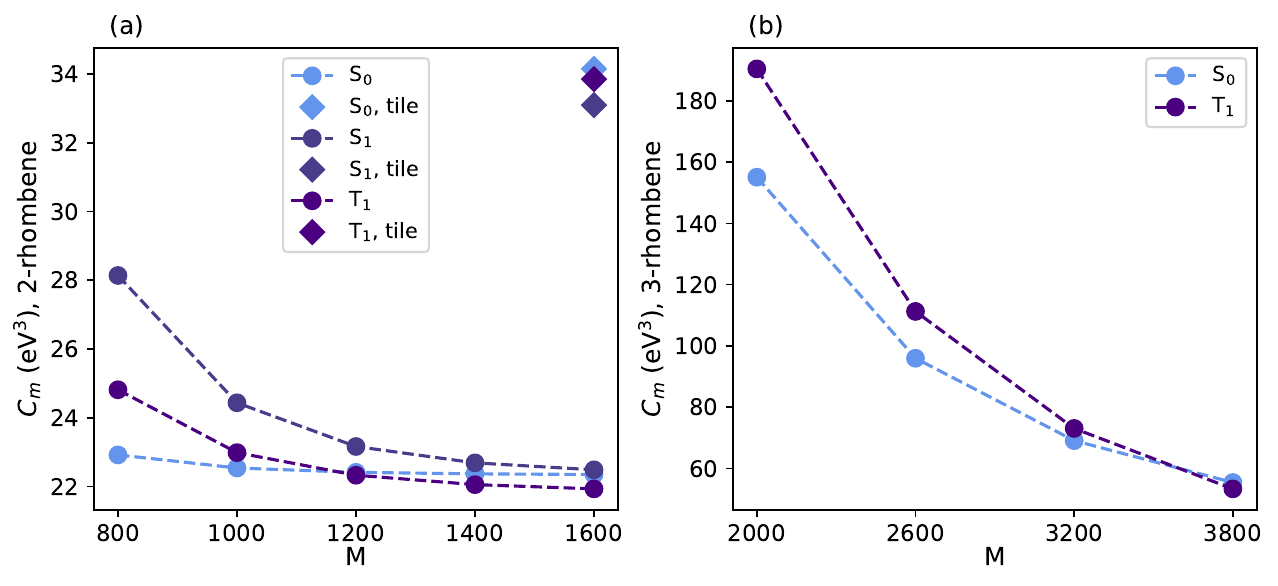}
    \caption{Energy error constant calculations of (a) 2-rhombene and (b) 3-rhombene eigenstates using our TD-DMRG method with bond dimension $M$. In (a), we show the energy error constants of the eigenstates $\mathrm{S_0}$, $\mathrm{S_1}$ and $\mathrm{T_1}$ of both $U_{\mathrm{SO}}$ and $U_{\mathrm{tile}}$. In (b), we only include the $\mathrm{S_0}$ and $\mathrm{T_1}$ energy error constants of $U_{\mathrm{SO}}$ -- the $\mathrm{S_1}$ state as well as $U_{\mathrm{tile}}$ calculations required substantial computational resources, and given that it would likely not provide additional insights, we chose not to perform these calculations.  The 3-rhombene $U_{\mathrm{SO}}$ energy error calculations shown in this plot are far from converged with respect to the TD-DMRG bond dimension, $M$.}
    \label{fig:C_estimates_rhombenes}
\end{figure*}

\begin{figure*}[!t]
    \centering
    \includegraphics[width=0.64\linewidth]{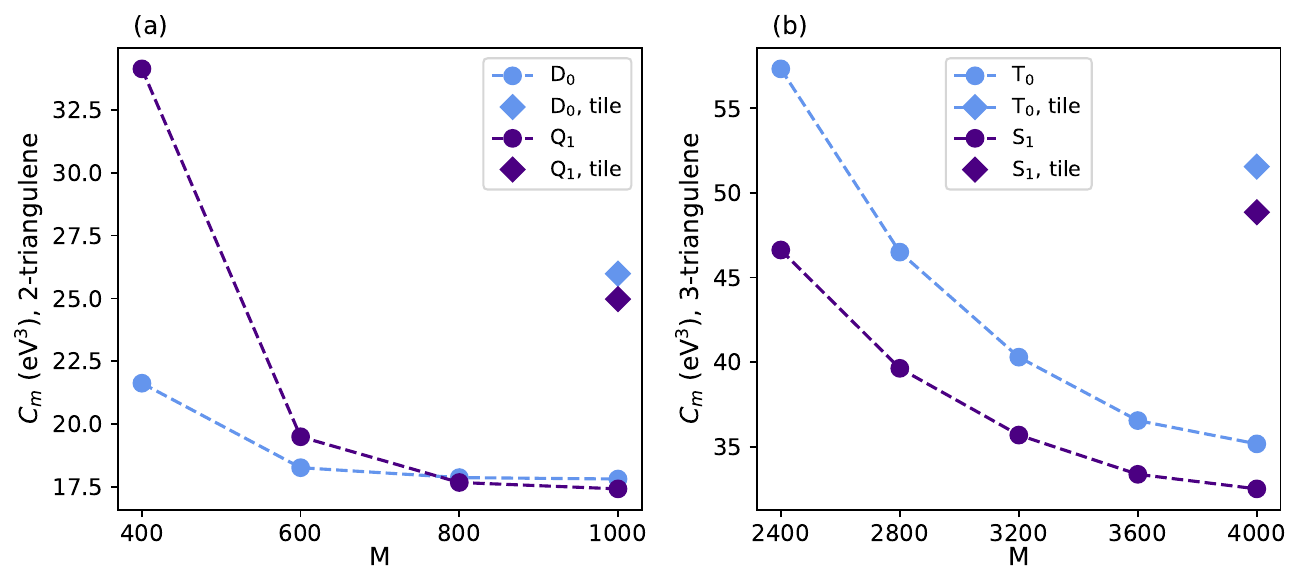}
    \caption{Energy error constant calculations of (a) 2-triangulene and (b) 3-triangulene eigenstates using our TD-DMRG method with bond dimension $M$. In (a), we show the energy error constants of the eigenstates $\mathrm{D_0}$ and $\mathrm{Q_1}$ (doublet ground state and a quartet excited state) of both $U_{\mathrm{SO}}$ and $U_{\mathrm{tile}}$. In (b), we show results for $U_{\mathrm{SO}}$ and $U_\mathrm{tile}$ of the triplet ground state $\mathrm{T_0}$ and the singlet excited state $\mathrm{S_1}$. While the 3-triangulene error constant calculations are not fully converged with respect to $M$, we expect the results obtained at $M=4000$ to be close enough to convergence to provide accurate estimates of effective energies and energy errors.}
    \label{fig:C_estimates_triangulenes}
\end{figure*}

\subsubsection{Time series analysis: results}
Our objective is to quantify energy and gap Trotter errors for relevant eigenstates of the nanographenes considered in this paper. For the acenes and rhombenes, we focus on the spin-states $m \in \{ \mathrm{S_0}, \mathrm{S_1}, \mathrm{T_1}\}$, where $\mathrm{S_0}$ is the singlet ground state, $\mathrm{S_1}$ is first-excited singlet and $\mathrm{T_1}$ is the first-excited triplet. Our time series analysis is computationally efficient for the quasi-1D acenes but becomes significantly more expensive when considering 2D systems. Therefore, we restrict ourselves to $n$-rhombene and $n$-triangulene with $n\leq3$. For the triangulenes, we consider the following spin eigenstates: 2-triangulene: $\mathrm{D_0}$ (ground-state doublet) and $\mathrm{Q_1}$ (first-excited quartet), and 3-triangulene: $\mathrm{T_0}$ (ground-state triplet) and $\mathrm{S_1}$ (first-excited singlet).

The DMRG and TD-DMRG calculations are performed using the Block2 code \cite{Zhai2023Block2}. We perform spin-adapted DMRG and TD-DMRG calculations (SU2-symmetry), and optimize the orbital-ordering of the MPS using the ``gaopt''-reordering algorithm. We use the time-dependent variational principle (TDVP) algorithm implemented in Block2 as our time-evolution algorithm and fix a maximum bond dimension, $M$, for the TD-DMRG calculations. We specify the maximum bond dimension used to obtain the energy and gap error results in this paper.

The TD-DMRG time evolution algorithm will also have an intrinsic time step error, which (similarly to Trotter error) can be controlled by decreasing the time step size. We monitored convergence of the effective energies with respect to the size of the TD-DMRG time-step and found that time steps up to $t = 0.25 \; \mathrm{eV}^{-1}$ in the TD-DMRG implementations of the 2nd order Trotter steps yielded sufficiently accurate effective energies and effective energy gaps. In practice, when implementing second order Trotter steps, the time step size for the time evolution of each individual part of $H$ is halved. While this is required to correctly implement second order Trotter steps, it also makes the accuracy less sensitive to TD-DMRG time step errors.

In Figs.~\ref{fig:C_estimates_acenes}, \ref{fig:C_estimates_rhombenes} and \ref{fig:C_estimates_triangulenes} we present energy error results using our time series analysis for the $n$-acene, $n$-rhombene and $n$-triangulene PPP models, respectively. Note that the result for $3$-rhombene and $3$-triangulene uses a large TD-DMRG bond dimension, $M$, and that the energy Trotter error constants of 3-triangulene and 3-rhombene are not fully converged with respect to $M$.

\begin{figure*}[!t]
    \centering
    \includegraphics[width=1.0\linewidth]{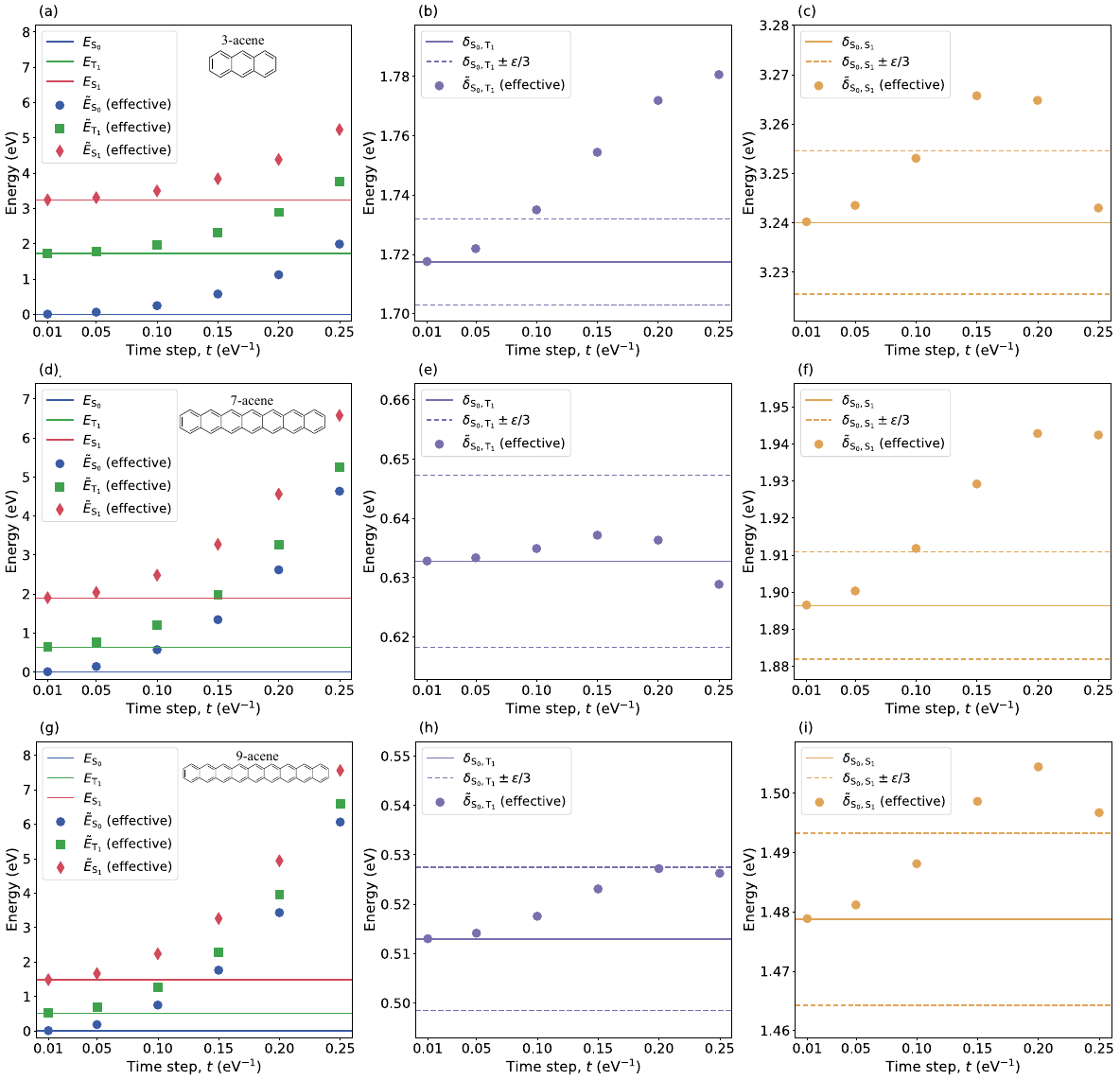}
    \caption{Exact energies, effective energies, exact energy gaps and effective energy gaps of (a), (b), (c) 3-acene, (d), (e), (f) 7-acene, (g), (h), (i) 9-acene using $U_{\mathrm{tile}}$ as our Hamiltonian simulation approximation. We refer to the caption of Fig.~\ref{fig:Trotter_relative_error_5_acene_main_text} in the main text for a detailed explanation.}
    \label{fig:realative_error_3_acene}
\end{figure*}

The exact energies, effective energies, exact energy gaps and effective energy gaps of $U_{\mathrm{tile}}$ presented in Fig.~\ref{fig:Trotter_relative_error_5_acene_main_text} in the main text and in Fig.~\ref{fig:realative_error_3_acene} are also calculated based on our TD-DMRG method and time-series analysis. We note in parentheses the maximum bond dimension ($M$) used for the TD-DMRG calculations of each system: Fig.~\ref{fig:Trotter_relative_error_5_acene_main_text} shows the energy and gap errors of 5-acene ($M=1200)$, 2-rhombene ($M = 1600$), 2-triangulene ($M=1200$), and 3-triangulene ($M=4000$), and Fig.~\ref{fig:realative_error_3_acene} shows the energy and gap errors of 3-acene ($M=600$), 7-acene ($M = 1400$) and 9-acene ($M=1500$).

\begin{table*}
\resizebox{\textwidth}{!}
{\begin{tabular}{lrrrrrrrrrrrrrr}
\hline
\hline
& \multicolumn{8}{c}{$n$-acene} & & \multicolumn{2}{c}{$n$-rhombene} & & \multicolumn{2}{c}{$n$-triangulene$^*$} \\
\cline{2-9}
\cline{11-12}
\cline{14-15}
 & 2 & 3 & 4 & 5 & 6 & 7 & 8 & 9 & $\;$ & 2 & 3 & $\;$ & 2 & 3  \\
\hline
$\delta_{\mathrm{S_0},\mathrm{T_1}}$ ($\delta_{\mathrm{gs},\mathrm{es_1}}$)$^*$ & \; 2.529  & \; 1.717 & \; 1.224 & \; 0.927 &  \; 0.745 & \; 0.632 &  \; 0.560 &  \; 0.513 & &  \; 1.875 & \; 0.645  & &  \; 3.590 &  \; 0.538 \\
\hline
$\delta_{\mathrm{S_0},\mathrm{S_1}}$ &  \; 3.611 &  \; 3.240 &  \; 3.045 &  \; 2.605 &  \; 2.204 & \; 1.896 &  \; 1.660 &  \; 1.479 & &  \; 3.070 & \; --  & &  \; -- &  \; -- \\
\hline
\end{tabular}}
\caption{Exact energy gaps between low-lying eigenstates of $n$-acene, $n$-rhombene and $n$-triangulene within the PPP model. All energy gaps are given in $\mathrm{eV}$ and rounded to the 3rd decimal.}
\label{tab:exact_energy_estimates}
\end{table*}

We make all time-series data files used to extract the energies and energy gaps of the effective Hamiltonians available on Zenodo (see data availability statement). Note that extracting the effective energies from the time-series analysis data will yield energies of a PPP Hamiltonian shifted by a constant energy chosen to symmetrize the spectrum around $E=0$ (to minimize wrapping of states within the entire spectrum). For an easier comparison to the exact energies, we therefore specify the converged exact eigenenergy gaps in Table~\ref{tab:exact_energy_estimates} (obtained from DMRG calculations) between the ground-state and excited state energies of interest of the nanographenes considered in this paper (disregarding the larger, computationally expensive 2D systems).

\section{Quantum phase estimation resource estimates \label{app:phase_estimation_costing}}
In our resource estimates, we disregard the cost of initial state preparation, although this is an important and non-trivial part of QPE \cite{Lee2023quantumadvantage}. We allocate $(1-x)\varepsilon$ error to Trotter and phase error, and allocate the remaining $x \varepsilon$ to rotation synthesis error \cite{Campbell_2022}, choosing $x = 0.02$ for all resource estimates, except for the Hamming weight phasing estimates in Fig.~\ref{fig:QPE_resource_estimates_HWP}, where we use $x = 0.025$. We use repeat-until-success (RUS) for synthesis of arbitrary rotations \cite{Bocharov2015EfficientCircuits}, which requires an additional qubit. The tile Trotter unitary requires a number of arbitrary rotations, $\NR$, and a number of additional T gates, $\NT$, per Trotter step. In Table~\ref{tab:per_trotter_step_costs}, we provide $\NR$ and $\NT$ for the PPP model Trotter steps of all systems for which we provide full non-Clifford QPE resource estimates. We provide the cost for implementing time-evolution of the potential energy operator and the kinetic energy operator (using the tile Trotterization splitting) separately. For sufficiently large number of Trotter steps, $N_{\mathrm{steps}}$, summing the cost for implementing $e^{-iVt}$ and $U_T$ (the tile Trotterization approximation of $e^{-iTt}$, see Eq.~\eqref{eq:U_T}) will give the full cost of a $U_{\mathrm{tile}}$ step, since
\begin{equation}
    U_{\mathrm{tile}}^{N_{\mathrm{steps}}} = e^{-iVt/2} \big( U_T e^{-iVt} \big)^{N_{\mathrm{steps}}} e^{iVt/2}.
\end{equation}
The arbitrary rotation cost of $e^{-iVt}$ is given simply by the number of terms in the symmetry-shifted potential energy operator, and this time evolution operator requires zero additional T gates. The cost of $U_T$ is lower but more involved since this requires tilings of the kinetic energy operator of each system. Given tilings of a kinetic energy operator, the gate counts can be evaluated according to a procedure described in Ref.~\cite{Bay-Smidt2025}. Note, we provide code (see data availability statement) to obtain the cost of implementing $U_T$ for each molecule with the tiling employed in this paper.

We split the remaining QPE costing into two parts: 1) using worst-case, average-case and energy errors and 2) using gap errors.

\begin{table*}
\resizebox{\textwidth}{!}
{\begin{tabular}{lrrrrrrrrrrrrrrrr}
\hline
\hline
& \multicolumn{5}{c}{$n$-acene} & & \multicolumn{4}{c}{$n$-rhombene} & & \multicolumn{4}{c}{$n$-triangulene} \\
\cline{2-6}
\cline{8-11}
\cline{13-16}
 & 3 & 5 & 7 & 9 & 13 & $\; $ & 2 & 3 & 4 & 5 & $\;$ & 2 & 3 & 4 & 5  \\
\hline
$\NR$ ($V$) &  \; 290 &  \; 794 &  \; 1554 &  \; 2570 &  \; 5370 &  &  \; 384 &  \; 1522 &  4128 &  \; 8994 & &  \; 241 &  \; 778 &  \; 1869 &  \; 3778 \\
$\NT$ ($V$) & 0 & 0 & 0 & 0 & 0 & & 0 & 0 & 0 & 0 & & 0 & 0 & 0 & 0 \\
$\NR$ ($T$) & 52 & 84 & 116 & 148 & 212 & & 64 & 124  & 208 & 308 & & 52 & 92  & 136 & 192 \\
$\NT$ ($T$) & 104 & 168 & 232 & 296 & 424 & & 112 & 248 & 400 & 616 & & 88 & 168 & 272 & 384 \\
\hline
\hline
\end{tabular}}
\caption{The arbitrary rotation and T gate cost of implementing time evolution of $V$ and $T$ using $U_{\mathrm{tile}}$. The cost is dominated by $V$, since each arbitrary rotation typically requires 30--50 T gates to implement. The cost of implementing time evolution of $T$ is more involved and can be obtained from the respective tilings of the kinetic energy operator. We refer to Ref.~\cite{Bay-Smidt2025} for a detailed costing of the tile Trotterization of the kinetic energy operator. We provide code (see data availability statement) that generates the kinetic operator sections of each molecule presented here, and which furthermore counts the number of tiles within each section and uses this to generate the resource estimates for implementing time evolution of $T$ presented here.}
\label{tab:per_trotter_step_costs}
\end{table*}

\textbf{1) Worst-case, average-case and energy error:} We find the Trotter error constant and use this to evaluate the required number of Trotter steps, $N_{\mathrm{steps}}$, \cite{Kivlichan2020ImprovedTrotterization, Campbell_2022}
\begin{equation}
    N_{\mathrm{steps}} = 6.203 \frac{\sqrt{G}}{(1-x)^{3/2} \varepsilon^{3/2}}, \label{eq:N_PE_worst_avg_absolute}
\end{equation}
where $G \in \{W, A, C \}$ depending on whether we use worst-case, average-case or energy errors. The total number of T gates required per Trotter step, $N_{\mathrm{T}/\mathrm{step}}$ can be evaluated as
\begin{equation}
    N_{\mathrm{T}/\mathrm{step}} = \NR \big(1.15 \log_2 \Big(\frac{\NR \sqrt{3G}}{x \sqrt{1-x} \varepsilon^{3/2}} \Big) +9.2\big)+\NT .  \label{eq:N_T_per_step_worst_avg_absolute}
\end{equation}
With these ingredients, we can simply evaluate the total T gate cost for a quantum phase estimation run
\begin{equation}
    N_{\mathrm{T},\mathrm{total}} = N_{\mathrm{steps}} \times  N_{\mathrm{T}/\mathrm{step}}.\label{eq:N_T_total_worst_avg_absolute}
\end{equation}
We can easily convert this into a Toffoli gate count by using the $\ket{\mathrm{CCZ}} \rightarrow 2 \ket{\mathrm{T}}$ conversion of Ref.~\cite{Gidney2019efficientmagicstate}, resulting in the following total Toffoli cost
\begin{equation}
    N_{\mathrm{Toffoli,total}} = N_{\mathrm{steps}} \times  \frac{N_{\mathrm{T}/\mathrm{step}}}{2}, \label{eq:N_Toffoli_total_worst_avg_absolute}
\end{equation}
using the standard implementation of $U_{\mathrm{tile}}$ (without additional tricks such as Hamming weight phasing to reduce the number of non-Clifford gates).

\textbf{2) Gap error:} We fix the time step, $t$, based on the considerations discussed in Fig.~\ref{fig:Trotter_relative_error_5_acene_main_text}. Using a fixed time step and allocating $\frac{2}{3} (1-x)\varepsilon$ error to phase estimation\footnote{\scriptsize Since we use phase estimation to measure each eigenenergy separately, the total error must account for the uncertainty of both QPE runs when calculating energy differences. For absolute energies, we follow Ref.~\cite{Campbell_2022} and use an additive error budget, adding the RMS (root mean square) readout uncertainty to the Trotter and synthesis error bounds. For a gap estimated from two independent QPE runs with unbiased readout, the mean-square error is the sum of the two readout variances and the squared gap bias, where the magnitude of the bias is bounded by the sum of the Trotter error in the gap and the synthesis errors from both runs. For our error allocation, the additive bound on the gap error, $\sqrt{2}\,\varepsilon_{\rm PE}+\varepsilon/3+2x\varepsilon \approx 2.08\,\mathrm{mHa}$, exceeds $\varepsilon$, whereas the resulting bound on the gap RMS error (as used in Ref.~\cite{gunther2025phaseestimationpartiallyrandomized}) is below. The reported resources are therefore sufficient to estimate the gap with an RMS error below chemical accuracy. Applying the same RMS error to absolute energies at $x=0.02$, with fixed Trotter step and synthesis precision, would reduce the absolute-energy gate counts by about $30\%$.}, we require  \cite{Kivlichan2020ImprovedTrotterization}
\begin{equation}
    N_{\mathrm{steps}} = \frac{2.28 \pi}{2(1-x)\varepsilon t} \label{eq:N_PE_relative}
\end{equation}
total number of Trotter steps. The number of T gates per Trotter step can be evaluated as \cite{Campbell_2022}
\begin{equation}
N_{\mathrm{T}/\mathrm{step}} = \NR \big(1.15 \log_2 \Big(\frac{\NR}{x \varepsilon t} \Big) +9.2 \big)+\NT . \label{eq:N_T_step_relative}
\end{equation}
The total number of T gates can be found by multiplying Eq.~\eqref{eq:N_PE_relative} with Eq.~\eqref{eq:N_T_step_relative}.

\subsection{QPE with Hamming weight phasing \label{sec:QPE_with_HWP}}
Hamming weight phasing (HWP) reduces the number of arbitrary-angle rotations needed to implement a batch of equal-angle rotations, at the cost of additional qubits and Toffoli gates \cite{Gidney2018halvingcostof, Campbell_2022}. Our HWP-QPE resource estimates account for these additional qubits and gates.

For the kinetic evolution $e^{-iTt}$, same-angle rotations within tiles of the same color act on disjoint qubits and can be batched as described in Refs.~\cite{Campbell_2022, Bay-Smidt2025}. The potential evolution $e^{-iVt}$ also contains many same-angle rotations, but some of them share a qubit. For example, if site $i$ has three nearest neighbors $j_1, j_2, j_3$, the three rotations generated by $Z_i Z_{j_k}$ have the same angle but all act on qubit $i$. Such rotations could be combined into a single batch by first mapping their parities onto distinct qubits with CNOT gates. For simplicity, we do not do this and only batch rotations that act on disjoint qubits.

To estimate the parallelism available for HWP, we group the $ZZ$ rotations in $e^{-iVt}$ by angle. For a group of $m$ rotations, let $\Delta$ be the largest number acting on any single qubit. These $\Delta$ rotations require separate batches, so $m/\Delta$ is an upper bound on the average batch size. For our gate-count estimate, we assume $\Delta$ batches per group and cost each batch at $\lceil m/\Delta\rceil$ rotations. We allocate $N-1$ HWP ancilla qubits to support batches of up to $N$ rotations, the maximum number of pairwise disjoint $ZZ$ rotations on the $2N$ data qubits. For most angles, even the upper bound $m/\Delta$ is well below $N$, so these groups use only a fraction of the available batch capacity on average. Our estimates thus illustrate the potential benefit of HWP for Trotterization of PPP Hamiltonians under this simple batching assumption.

\end{document}